\documentclass[12pt,preprint]{aastex}

\usepackage{color}
\usepackage{graphicx}
 
\newcommand{\Lya}{\mbox{Ly$\alpha$}}
\newcommand{\lya}{\mbox{Ly$\alpha$}}
\newcommand{\lyb}{\mbox{Ly$\beta$}}

\newcommand{\kms}{\mbox{km s$^{-1}$}}
\newcommand{\cmm}{\mbox{cm$^{-2}$}}

\newcommand{\etal}{{\it et al.}}
\newcommand{\zabs}{\mbox{$z_{\rm abs}$}}

\newcommand{\lyaf} {\lya\ forest}
\newcommand{\Lyaf} {\lya\ forest}
\newcommand{\nhi}{\mbox{N$_{\rm H I}$}}
\newcommand{\lnhi}{\mbox{log \nhi}}

\newcommand{\zem}{\mbox{$z_{\rm em}$}}

\newcommand{\ol}{\mbox{$\Omega_{\Lambda }$}}
\newcommand{\ob}{\mbox{$\Omega_b$}}
\newcommand{\obh}{\mbox{$\Omega_bh^{2}$}}
\newcommand{\om}{\mbox{$\Omega_m$}}

\newcommand{\lamr}{\mbox{$\lambda_{\rm r}$}}
\newcommand{\lamo}{\mbox{$\lambda_{\rm o}$}}
\newcommand{\taueff}{\mbox{$\tau _{\rm eff}$}}
\newcommand{\sigzo}{\mbox{$\sigma (\Delta z = 0.1) $}}

\newif\ifdraftmodep
\draftmodepfalse

\newif\ifapjp
\apjpfalse

\newcommand{\dadef}[2]{#2}

\newcommand{\fmce}{\dadef{fmce}{3.5}}
\newcommand{\daZerosmp}{\dadef{daZerosm}{15.1}}
\newcommand{\daQErrorp}{\dadef{daQError}{0.7}}
\newcommand{\dasZerosp}{\dadef{dasZerosp}{20.3}}

\newcommand{\dasfonem}{\dadef{dasfonem}{0.1687}}
\newcommand{\dasftwom}{\dadef{dasftwom}{0.1585}}
\newcommand{\dasfthreem}{\dadef{dafthreem}{0.1602}}
\newcommand{\dastruem}{\dadef{dastruem}{0.1601}}

\newcommand{\darfonem}{\dadef{darfonem}{0.1637}}
\newcommand{\darftwom}{\dadef{darftwom}{0.1533}}
\newcommand{\darfthreem}{\dadef{darfthreem}{0.1552}}

\newcommand{\datwosp}{\dadef{datwosp}{6.81}}
\newcommand{\dastwosp}{\dadef{dastwosp}{6.12}}

\newcommand{\zfitgamma}{\dadef{zfitgamma}{2.57}}

\newcommand{\ftcl}{\dadef{ftcl}{0.970}}
\newcommand{\ftch}{\dadef{ftch}{1.051}}

\newcommand{\ftcs}{\dadef{ftcs}{5.4}}
\newcommand{\ftcms}{\dadef{ftcms}{1.7}}

\begin{document}

\title{
Cosmological parameters $\sigma_8$, the baryon density \ob, and 
the UV background intensity from a calibrated measurement of
H~I Lyman $\alpha$ absorption at $z = 1.9$
\altaffilmark{1}} 

\author{ 
David Tytler\altaffilmark{2},
David Kirkman,
John M. O'Meara,
Nao Suzuki,
Adam Orin,
Dan Lubin,
Pascal Paschos,
Tridivesh Jena,
Wen-Ching Lin,
\& Michael L. Norman \\
Center for Astrophysics and Space Sciences;\\
University of California, San Diego; \\
MS 0424; La Jolla; CA 92093-0424\\}

\altaffiltext{1} {Based on data obtained with the Kast spectrograph 
on the Lick Observatory 3-m Shane telescope.
}
\altaffiltext{2} {E-mail: tytler at ucsd.edu}

\begin{abstract}

We identify a concordant model for the intergalactic medium (IGM)
at redshift $z=1.9$ that uses popular values for cosmological 
and astrophysical parameters and accounts for all baryons 
with an uncertainty of 6\%. 
The amount of absorption by H~I in the IGM
provides the best evidence on the physical conditions in the IGM,
especially the combination of the mean gas density, the density
fluctuations, the intensity of the ionizing flux, and the level of
ionization.  We have measured the amount of absorption, known as the
flux decrement, DA, in the \Lyaf\ at redshift 1.9. We used spectra of
77 QSO that we obtained with 250~\kms\ resolution from the Kast
spectrograph on the Lick observatory 3m telescope.  We fit the
unabsorbed continua to these spectra using b-splines.  We also fit
equivalent continua to 77 artificial spectra that we made to match the
real spectra in all obvious ways: redshift, resolution, S/N, emission
lines and absorption lines.  The typical relative error in our
continuum fits to the artificial spectra is \fmce \%.  Averaged over
all 77 QSOs the mean level is within 1--2\% of the correct value,
except at S/N $< 6$ where we systematically placed the continuum too
high.  We then adjusted the continua on the real spectra to remove
this bias as a function of S/N and a second smaller bias.  Absorption
from all lines in the \lyaf\ at $z = 1.9$ removes DA(z=1.9) =
\daZerosmp $\pm$ \daQErrorp \%
of the flux at rest frame wavelengths $1070 < $ \lamr\ $ < 1170$~\AA .
This is the first measurement using many QSOs at this $z$, and the
first calibrated measurement at any redshift.  Using similar methods
on $1225 < $ \lamr\ $ < 1500$~\AA\ we find metal lines absorb an
average 1.6\% the flux, increasing slightly as the rest frame
wavelength \lamr\ decreases because more types of spectral lines
contribute and there is more C~IV at lower redshifts.  We estimate
that the metal lines absorb $2.3 \pm 0.5$\%
of the flux in the \lyaf\ at z=1.9. 
The absorption from \lya\ alone then has 
DA $= 12.8 \pm 0.9$\%. 
The \lya\ lines of Lyman limit systems with column densities \lnhi
$>17.2$~\cmm\ are responsible for a DA $=1.0 \pm 0.4$\% at
$z=1.9$. These lines arise in higher density regions than the bulk of
the IGM \lya\ absorption, and hence they are harder to simulate in the
huge boxes required to represent the large scale variations in the
IGM.  If we subtract these lines, for comparison with simulations of
the lower density bulk of the IGM, we are left with DA $= 11.8 \pm
1.0$\%.
The mean DA in segments of individual spectra with $\Delta z= 0.1$, or 
153~Mpc comoving at $z=1.9$, has a large dispersion, $\sigma =6.1\pm 0.3$\%
including Lyman limit systems (LLS) and metal lines,
and \sigzo $= 3.9 ^{+0.5}_{-0.7}$\% for the \lya\ from the 
lower density IGM alone, excluding LLS and metal lines.
This is consistent with the usual description of large scale structure 
and accounts for the large variations from QSO to QSO.
Although the absorption at $z=1.9$ is mostly from the lower density IGM, the
\lya\ of LLS and the metal lines are both major contributors to the 
variation in the mean flux on 153~Mpc scales 
at $z=1.9$, and they make the
flux field significantly different from a random Gaussian field 
with an enhanced probability of a large amount of absorption.
We find that a hydrodynamic simulation on a $1024^3$ grid in a 75.7~Mpc box 
reproduces the observed DA from the low density IGM alone when we use 
popular parameters values H$\rm _o=71$~\kms Mpc$^{-1}$,
$\Omega_b = 0.044$, $\Omega_m = 0.27$, ${\Omega_{\Lambda} = 0.73}$,
${\sigma_8 = 0.9}$ and a UV background (UVB) that has an ionization rate per H~I atom
of $\Gamma _{912}= (1.44 \pm 0.11) \times 10^{-12}$~s$^{-1}$. This is 
$1.08 \pm 0.08$ times the prediction by Madau, Haardt \& Rees (1999)
with 61\% from QSOs and 39\% from stars. 


\end{abstract}

\section{Introduction}
\label{sa}

Our physical understanding of the IGM comes from the detailed
comparison of numerical simulations of the growth of structure in the
universe with observations of the \lya\ absorption from the H~I in the
IGM.  The amount of \lya\ absorption depends on a combination of at
least four factors: the mean density of H in the IGM, the power
spectrum of the matter distribution that determines the amount of
clumping of the H on various scales, the temperature of the gas, and
especially the mean intensity of the UVB radiation
that photoionizes the gas.  Together these parameters, and their
variation, give the density of H~I down the line of sight to a QSO,
something we observe with \lya\ absorption.

The mean amount of absorption is a sensitive measure of the physical
properties of the IGM. If we make some assumptions about the growth of
structure and the temperature of the IGM, then the optical depth of
the \lyaf\ scales like \citep[Eqn. 17]{rauch97}
\begin{equation}
\tau_{{\rm Ly} \alpha} \propto (1+z)^6 H(z)^{-1}
(\Omega_b h^2)^2 T^{-0.7} (\rho / < \rho > )^{\alpha } \Gamma _{912}^{-1},
\end{equation}
where $\rho $ is the proper baryon density, $\Gamma _{912}$ is the
photoionization rate per H~I atom, and $T$ is the gas temperature. The
exponent $\alpha = 2$ for isothermal gas, and it is 1.6 -- 1.8 for the
low density IGM, because denser gas is hotter
\citep{hui97,croft97a,croft02a}.  To first order, the amount of
absorption at a given wavelength in a spectrum reflects the density of
H~I in part of the IGM, which comes from the density of gas and dark
matter.

Many authors have to used this relationship, together with an estimate
of $\Gamma _{912}$, to estimate the cosmological baryon density from
the mean amount of absorption in the \lyaf. The \obh\ estimates have
all been too large \citep[Fig. 4]{haehnelt01}.  \citet{rauch97}
estimated $\Gamma _{912} > 7 \times 10^{-13}$ s$^{-1}$ due to the
contribution to the UVB from known QSOs, which is consistent with the
$\Gamma _{912} = 8.15 \times 10^{-13}$ s$^{-1}$ due to QSOs from
\citet{haardt96}.  This corresponds to $\Omega_b h^2 > 0.021$ in a
$\Lambda$CDM model.  \citet{steidel01} found that a large fraction of
ionizing photons escape from Lyman break galaxies, giving $\Gamma
_{912}> 1.5 \times 10^{-12}$ s$^{-1}$. \citet{haehnelt01} used this to
conclude $\Omega_b h^2 > 0.06$ in their $\Lambda$CDM models, while
\citet{hui02} assumed different mean temperature for the IGM and found
$\Omega_b h^2 = 0.045 \pm 0.008$.  These values are higher than the
more robust measurements of $\Omega_b h^2 = 0.021\pm 0.002$ from our
measurements of D/H using Standard Big Bang Nucleosynthesis
\citep{kirkman03}, and $\Omega_b h^2 = 0.0224\pm 0.0009$ from the
anisotropy of the CMB \citep{spergel03}.  In this paper we present a
more accurate measurement of the mean amount of absorption in the
\lyaf, DA, and for the first time we find that this is consistent with
\ob $= 0.044$ in a popular cosmological model.

An accurate measurement of DA is also a critical input to the
measurement of the power spectrum of matter using the \lyaf\
\citep{croft02b}. This is because a smaller DA requires a larger
amplitude for the matter power spectrum, if all other factors are
unchanged.  The larger the matter power amplitude, the less gas is
left widely distributed in the IGM where it causes the most
absorption.  For a given simulation, DA determines the relationship
between the mass field and the \Lyaf\ optical depth -- which we use to
infer the \lyaf\ mass power spectrum from the \lyaf\ flux power
spectrum. Hence, DA can be used to fix the constant of proportionality
in Equation 1 above. DA also has an effect on the shape of the power
spectrum deduced from the \lyaf\ (\citealt{zaldarriaga03},
\citealp[Figure 1b]{seljak03}).

Whether the power spectrum measured by the \lyaf\ might have lower
amplitude that that expected from measurements on larger scales from
galaxies and the CMB \citep{spergel03} depends upon the accuracy of
the \lyaf\ DA measurements.  \citet{croft02b} stressed that the
uncertainty in DA is the main source of error in estimates of the
matter power spectrum from the \lyaf .  Their Figure 17 shows how the
amplitude of the matter power spectrum changes with the mean opacity,
for a fixed flux power.  They chose \taueff $(z=2.72)= 0.349$ from
\citet{press93}.  If instead we now choose \taueff $ (z=2.72) =
0.280$, from the right hand of Figure 1 of \citet{schaye03a}, which we
expect is more accurate, the \citet{croft02b} results show that the
matter power spectrum rises by a factor of 2.1, assuming no change to
the temperature-density relation in the IGM.  \citet{seljak03} also
discussed this situation in detail. They reviewed DA estimates and
chose \taueff $(z=2.72) = 0.298$, from \citet{mcdonald00a}.  They too
find that this change increases the matter power by a factor of two
compared to the value in \citet{croft02b}, when combined with an
increase in the slope of the matter power spectrum.

\citet[Eqn. 12]{croft02b} find that the 3D matter power spectrum
amplitude $\propto \tau_{eff} ^{-3.4}$, for a given observed flux
power.  This implies that to measure the power spectrum amplitude to
10\% at this $z$, we would need a relative error on \taueff\ of 3\%,
which is a relative error on DA of 2.4\%, assuming that all other
factors were well known, which is not the case.  We find a similar
scaling relation from \citet{seljak03}.

However \citet[Eqn. 8]{gnedin02} find a much weaker correlation
between the matter power and DA. Although the origin of this
disagreement is unknown, \citet{seljak03} suggest that
\citet{gnedin02} did not sample a wide enough range of parameters.
Larger amplitude power comes with higher velocities which decreases
the flux power on small scales. As the matter power increases, the
flux power can both rise on large scales and fall on small scales.

These results show that high accuracy DA measurements are of great
interest because of their cosmological significance.  In general,
accurate measurements of DA can become a cornerstone in a concordance
model of the IGM, tying together the intensity of the UVB, the thermal
history of the IGM and the cosmological matter power spectrum.

\subsection{ Definition of DA}

Following \citet{oke82} we define DA = $1- \left< F\right> $ where
$\left< F\right> =$ (observed flux)$/C$, $C =$ (estimated unabsorbed
continuum flux) which includes both the underlying power law and the
flux from emission lines.  It is common to see DA expressed as $\left<
F\right>$, or as the mean effective optical depth, \taueff $ = -ln
\left< F\right>$.

We will measure DA in individual pixels, and we will use suffixes on
the DA to label averages over various wavelengths between the \lya\
and \lyb\ emission lines, sometimes averaged over many QSOs.  We
restrict our measurement of DA to rest frame wavelengths
\begin{equation}
\rm DA~ wavelength~ range~ = 1070 ~to~ 1170~\AA .
\end{equation}
The DA is dominated by \lya\ lines from the IGM, but it includes all
absorption, including metal lines and the \lya\ lines from 
LLS that by definition include all damped \lya\ lines
(DLAs).

After we measure DA, we will give estimates for the amount of
absorption due to the \lya\ lines of high column density absorption
systems and metal lines, both of which are harder to simulate.

\subsection{ Prior Measurements of DA}

More than a dozen papers contain measurements of DA: see references in
\citet{rauch98} and \citet{bernardi03} and the discussion of errors in
\citet{croft02b} and \citet{seljak03}.

DA is hard to measure because we need many QSO spectra, the unabsorbed
continuum level is hard to estimate, and when we want just the H~I
portion of the DA, the metal lines in the \lyaf\ are difficult to find
and measure.

To measure the mean DA with a relative error of 1\%, i.e. DA $= 0.300
\pm 0.003$, in a specific redshift range, we must observe about 10,000
\lya\ lines with \nhi\ values similar to those that makes most of the
optical depth \citep[Fig. 8]{kirkman97a}.  We need of order 500 QSOs
at $z=3$.  Most samples have used under ten QSOs.

The unabsorbed continuum is relatively easy to find in high resolution
spectra with high S/N, but we have few of these spectra. Instead, the
continuum in the \lyaf\ has often been set to a power law extrapolated
from wavelengths $> 1250$~\AA , and this can be biased
\citep{kim01,meiksin01b, seljak03}.

Most papers have not attempted to find and remove metal lines. Instead
they include such lines in their DA values.

These difficulties have lead to large differences in reported DA
values.  \cite{jenkins91} noted that the distribution of flux in the
\lyaf\ region implies that there is about 30\% more absorption than is
expected from lines identified in spectra, which is a huge
uncertainty.  Meiksin (1997, private communication) long ago drew our
attention to these disagreements. For example, at $z \simeq 3 $ we
have \taueff $ = 0.28$, 0.38 and 0.45 from \citet{hu95,rauch97} and
\citet{press93} respectively, while at $z=2.72$ we have \taueff $ =
0.28$ from \citet{schaye03a} and 0.35 from both \citet{press93} and
\citet{bernardi03}.

The uncertainty over DA has hindered attempts to measure parameters
such as \ob\ from the \lyaf .  Zhang et al. (1998)\nocite{zhang98}
obtained DA from a conventional power law fit to the \nhi\
distribution fit from Hu et al. (1995)\nocite{hu95}, who in turn had
used Keck spectra of 4 QSOs all at $z = 3.1 - 3.4$. We know that the
integral over this power law could contain large errors. Weinberg et
al. (1997)\nocite{weinberg97} used Press, Rybicki \& Schneider
(1993)\nocite{press93} very low resolution (25\AA) spectra of 29 QSOs
at $z = 2.5 - 4.3$.  We will discuss the most detailed study, by
\citet{rauch97} later.

Two recent measurements of DA are of special interest.  Bernardi et al
(2003)\nocite{bernardi03} used 1061 SDSS spectra to measure DA at $2.6
<$ \zabs\ $< 4.0$, again outside our range.  They introduce new
methods and obtain by far the best random error although there seem to
be systematic problems with their results.  \cite{schaye03a} used 21
UVES and HIRES spectra, with S/N $>40$ and they removed metal lines.
They found \taueff\ values that are systematically lower than
\citet{bernardi03} by 0.1 dex.

\subsection{What we will do}

Using artificial spectra with similar characteristics to the Kast
spectra set (e.g. S/N, resolution, redshift, and continuum shape), we
will characterize and statistically account for errors such as
continuum placement. We will aim for an absolute error in the mean DA
value of $<1$ \%, dominated by the random noise coming from the sample
size.

The paper is organized as follows: in \S \ref{sb} to \S \ref{se} we
describe our data set and calibration methods.  In \S \ref{sb} we
describe the Kast spectra; in \S \ref{sc} we describe the artificial
spectra that we created to calibrate the continuum fit; in \S \ref{sd}
we describe the continuum fits; and in \S \ref{se} we measure and
correct the errors we made in the continuum fits.

In \S \ref{sf} to \S \ref {si} we describe the results of our DA
measurement.  In \S \ref{sf} we give our measurement of DA. In \S
\ref{sg} we describe the dispersion we see in DA, and in \S
\ref{error} we summarize the error on DA.  In \S \ref{si} we summarize
and discuss our results.

To convert from redshift to distance, we use a Hubble constant H$\rm
_o=71$~\kms Mpc$^{-1}$, a vacuum energy of \ol $= 0.73$, and a matter
density of \om $= 0.27$ and we evaluate at $z=1.9$.

\section{Kast QSO Spectra}
\label{sb}

We obtained spectra of the \lyaf\ of bright QSOs at $1.85 <$ \zem\ $ <
2.5$. We maintained a list of all such QSOs listed in NED, and updated
it before each observing run. We found about 6000 QSOs of all
magnitudes. We rejected those noted as BAL in NED, and we then
observed the brightest remaining at declination North of $ -30$
degrees. Most were 17th magnitude. We observed nearly all that were
brighter than 17.5 and some that were 18th magnitude.  We rejected BAL
QSOs because they tend to show much more absorption than other
QSOs. In the \lyaf\ region this absorption is from N~V, and other
ions. BAL QSOs would also bias our estimates of the mean amount of
metal line absorption because they can have huge amounts of C~IV and
Si~IV absorption.

\subsection{Observations}

We obtained spectra from 2001 January 26 to 2003 July 28 with the Kast
double spectrograph on the Shane 3m telescope at Lick
observatory. Here we discuss spectra obtained with the blue camera,
using the grism with 830 groves per mm, blazed at 3460~\AA , and
covering approximately 3150 -- 4300~\AA .  We also took simultaneous
spectra with the red camera with the 1200 grove/mm grating blazed at
5000~\AA , giving 1.17~\AA\ /pixel and covering 4400~\AA\ to the red
of C~IV emission. We will not discuss those red spectra here.

We used the slit that is 2 arcsec wide when the seeing was good, but a
3 arcsec or wider slit was sometimes necessary.  The slit was aligned
with the vertical direction on the sky in the middle of each exposure.

We know from prior work with this instrument \citep{suzuki03b} that
the typical dispersion is 1.13~\AA\ per /pixel (107~\kms ), and the
FWHM resolution is 250~\kms\ (2.5 pixels), with a range of 200 --
300~\kms , depending on the temperature and the focus that we chose
for that observing run.  The spectral resolution varies with
wavelength, and from run to run, even when the slit is unchanged.

\subsection{Reductions}

We extracted and reduced the spectra with the standard IRAF long slit
reduction packages.  We performed the wavelength and flux calibrations
in the standard manner, again using IRAF. We gave a detailed
discussion of similar reductions of spectra from the same instrumental
setup in \citet{suzuki03b}.  Wavelength errors are typically under
1~\AA .  The standard extinction correction was applied, but we have
not performed a separate correction for the fluctuations in the ozone
absorption.  We calibrated the flux using Kast spectra (taken on the
same night as the QSO spectrum) of one or more of the following flux
standard stars: 
BD+28 4211, 
BD+33 2642, 
Feige34, 
Feige67 and
G191b2b. 
We took the fluxes for these stars from HST spectra that we have shown
are ideal for the wavelengths of interest \citep{suzuki03b}.  In a few
cases we found that calibrations using different stars differed.  

We changed the flux in the occasional pixel that was clearly erroneous
because of poor cosmic ray or sky subtraction. We set such pixels to
the expected level, to reduce the effect on the continuum fitting and
the DA estimate. Hence the spectra are cosmetically unusually clean.

\subsection{QSO Sample used}

We attempted to set the integration times to reach S/N = 10 per pixel
at 3200~\AA , although weather sometimes prevented us from achieving
this.  The S/N in the continuum of the spectra from \lyb\ to \lya\
vary from a few to over 50, with typical values of 6 -- 20.  Some of
this variation is from QSO to QSO, at a given wavelength, and some is
variation with wavelength within each spectrum.  For all spectra, the
S/N increases systematically with wavelength, and hence with \zabs .
This has important consequences that we discuss below.

We rejected about 10 QSOs because the SNR was $<2$. We rejected
Q2310+0018 (RA 23h10m50.80s +00d18m26.4s B1950, \zem = 2.200
mag. 17.00) because we discovered that it shows BAL absorption.  The
spectra that we obtained for 6 objects were not QSOs. Two were
probably our error in pointing the telescope. It is possible that
several of the other four are not QSOs:
Q1456+5404, 14h56m47.71 +54d04m25.6 \zem = 2.300  16.50mag; 
Q1742+3749, 17h42m 5.55 +37d49m08.3 \zem = 1.958  16.40mag;
Q1755+5749, 17h55m15.97 +57d49m06.9  \zem = 2.110  18.00mag; and
Q2113+3004, 21h13m59.42 +30d04m02.4  \zem = 2.080  17.30mag.

After these various rejections we are left with \lyaf\ spectra of 77
QSOs, which we use in this paper.  We measured the \zem\
for each QSO from the emission lines in its blue spectrum, typically
\lya , Si~IV and C~IV.

\begin{figure}
\epsscale{0.7}
\plotone{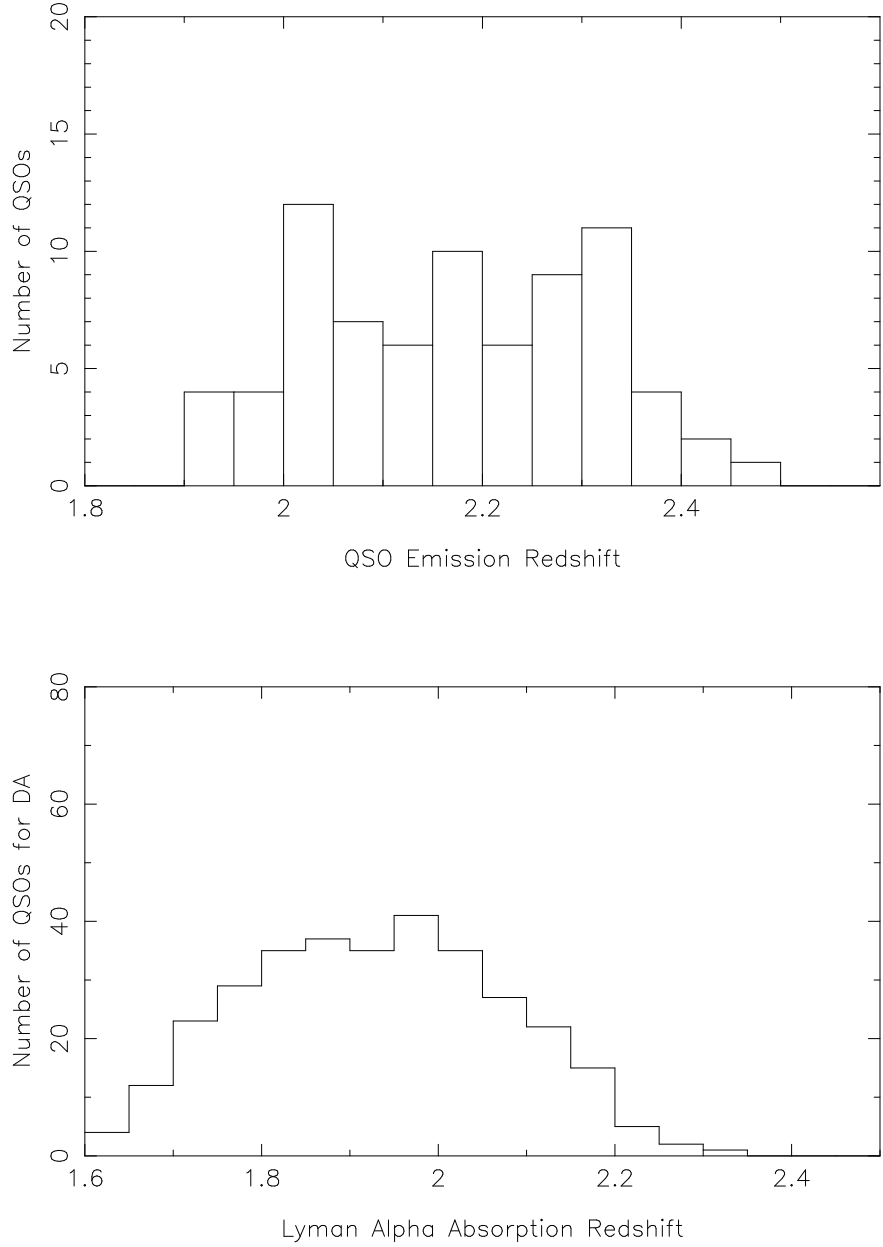}
\caption{\label{fx}
{\bf Top panel:} The number of QSOs in our sample as a function of
emission redshift. {\bf Bottom panel:} The number of QSOs which
contribute to our measurement of DA at each redshift.}
\end{figure}

In Figure \ref{fx} we show a histogram of the \zem\ values, and the number
QSOs contributing \lyaf\ information at various \zabs\ values.  The
mean \zem $= 2.17$ while the mean \zabs $= 1.924$.  Had we weighted by
S/N, the mean \zabs\ would have been $>2$.  We use spectra with
observed wavelengths 3173 -- 4083~\AA\ with a range of $\pm 5$~\AA\
from the precise CCD placement. This corresponds to \zabs\ = 1.6105 --
2.3587 for \lya.  Eight of the QSOs with \zem $<1.965$ do not cover
the whole 1070 -- 1170~\AA\ range, since their 1070 is $<3173$~\AA .

When we use the entire sample to measure DA, which we will call DA4,
we sample a total redshift path of 19.750 (30.8 comoving Gpc) in
23,000 pixels of 1.13~\AA\ in the observed frame.  At a redshift of
1.9, one pixel corresponds to 1.56 comoving Mpc, for a model with
H$\rm _o=71$~\kms /Mpc, \ol $= 0.73$, and \om $= 0.27$.

\section{Artificial QSO Spectra}
\label{sc}

We have made a set of 77 artificial spectra, each one matched to one
of the Kast spectra.  Each artificial spectrum has the same \zem\ and
S/N distribution as one of the Kast spectra.  All the artificial
spectra have shapes, including emission lines, from real HST spectra
of lower redshift QSOs, they all have random \lya\ absorption lines
from simulations of the IGM, and they have the same spectral
resolution as the Kast spectra.

The starting point for the artificial spectra were the smoothed
absorption free fits to the continuum and emission lines of the 50
QSOs discussed and listed in Table 1 of \citet{suzuki04a}.  They have
$0.14 < $\zem\ $<1.04$ and an average S/N = 19.5 per 0.5~\AA\ in the
rest frame from 1050 -- 1170~\AA .  The HST continua on these spectra
had previously been adjusted to match our understanding of QSO
continua. We believe that the continua levels are better known that
those for the Kast spectra, because the S/N is relatively high and
there are far fewer absorption lines at these low redshifts.  In a quick
look, the shapes of these HST spectra are not obviously different from
those of the QSOs that we observed with the Kast spectrograph.

We randomly associated each of the HST spectra with one of the Kast
spectra, and we use 27 of the HST spectra twice. These associations
match each of the \zem\ values to one and only one of the artificial
spectra.  We trimmed wavelength range of each artificial spectrum to
match the range of its paired Kast spectrum.

Next, we added \lya\ absorption from heuristic simulations of the
\lyaf .  The model used for the forest is a simplified version of the
\citet{bi92} log-normal model, which incorporates non-linear effects
into the linear theory of structure formation.  In the log-normal
model, the transmission fraction field is simply defined to be
\begin{equation}
F(\lambda) = \exp\left(-\tau_0
\exp\left[\delta\left(\lambda\right)-\sigma^2/2\right]^2\right)~,
\end{equation}
where $\delta(\lambda)$ is a Gaussian random field.  The power
spectrum of $\delta$, $P_\delta(k)$, was constructed to make the
simulated power spectrum of $F$, $P_F(k)$, roughly match the observed
$P_F(k)$ (e.g., from \citet{mcdonald00a}).  We made $\tau_0(z)$ and
the amplitude of $P_\delta(k,z)$ slowly varying functions of redshift
to match observations.  The change in the total absorption due to
\lya\ in these simulated spectra should follow $A(1+z)^\gamma$, where
$A=0.0166$ and $\gamma = 2.07$.  The simulations were made with full
numerical resolution, and then smoothed to a FWHM of 250~\kms .

We added the absorption starting at 1215.67~\AA\ (rest frame, as are the 
rest of the wavelengths in this section).   This is typically
near the peak of the \lya\ emission line using the \zem\ value that we
had measured. The absorption was added by multiplying the smoothly
changing fit to the HST spectrum by flux values from 0 -- 1. We
continued the absorption down to the UV end of the spectrum. The
spectra contain only \lya\ absorption, even when we are at wavelengths
where \lyb\ would also appear in Kast spectra. We did not make any
adjustments for the proximity effect, and hence the absorption from
the \lyaf\ in the artificial spectra begins at exactly 1215.67~\AA .

These simulations produce \lyaf\ absorption that is similar to Kast
spectra, but they were not adjusted to be as close as possible.
Compared to Kast spectra, the simulations have too few \lya\ with
large equivalent widths coming from the LLS, and they have no DLAs.
Otherwise, by visual inspection alone, we can not tell the artificial
spectra from the real \lyaf .  The lack of a proximity effect region
is not a distinction, since this can not be seen in a single spectrum.

We added a Gaussian random deviate to each pixel in each artificial
spectrum to make its S/N similar to that of its partner.  This
procedure is complicated because we wish to simulate the S/N that we
would have obtained with Kast, had we observed the QSO with the \zem\
of its partner and the spectral slope and emission lines of the HST
spectrum. Hence we can not simply copy the S/N from the Kast spectrum
partners, because the emission lines differ.

The S/N that we gave to an artificial spectra had the same values as
the S/N in its partner spectrum at two reference wavelengths, 1100 and
1255~\AA , and it responds to emission lines according to the $\sqrt
{flux}$ in the artificial spectrum.

In detail we multiplied the flux in each artificial spectrum by the
response function of the Kast, to simulate the distribution of photons
that we would have recorded, had we observed the HST flux
distribution.  The square root of this gives the relative S/N as a
function of wavelength.  We then measured the S/N in both the
artificial and Kast spectra averaging over 0.02 in z (24.3~\AA\ rest)
around each reference wavelength.  We took the ratios of these S/N
values, Kast upon artificial, to derive correction factors. We fit a
straight line between the two correction factors and then multiplied
the S/N on the artificial spectrum by this line.  The artificial and
Kast spectrum then have the same S/N at the reference wavelengths, and
generally similar distributions of S/N with wavelength.  We checked
that the distribution of S/N in the DA region in the 77 artificial
spectra was very similar to that in the 77 Kast spectra.

In Figure \ref{fxx} we show two artificial spectra. The only easy way to see
that these are artificial is that there is zero absorption to the red
of \lya .

\section{Continuum fitting}
\label{sd}

The methods that have been used to estimate the unabsorbed continuum
in the \lyaf\ fall into two general classes:

\begin{itemize}
   \item Extrapolations from wavelengths $>1250$~\AA\ (rest frame, as are 
     the other wavelengths in this section) that do not use
     the flux information from the \lyaf . The extrapolations use
     power laws \citep{oke82, steidel87a, bernardi03}, similar smooth
     functions \citep{press93}, or principal components that also
     predict the shapes of emission lines in the \lyaf\
     \citep{suzuki04a}.
   \item Fits to the local continuum in the \lyaf\ that emphasize the
     wavelengths with the most flux
     \citep{rauch97,fang98,mcdonald00a,kim02b,schaye03a}.
\end{itemize}

The values that we quoted in \S \ref{sa} illustrate that the
extrapolations give systematically much more DA than do fits to the
local continuum, a point noted by others \citep{kim01,meiksin01b,
seljak03}. Like other authors, we suspect that the extrapolated
continua are less reliable, and too high in the \lyaf .

\citet{seljak03} suggest that the extrapolated continuum is too high
because QSO spectra are not well fit by a power law with a single
slope.  It is well established that best fit power law declines faster
with decreasing wavelength at $<1200$ than at $>1300$~\AA\
\citep[Figure 4]{telfer02}.  \citet{seljak03} calculate that the DA
from power law extrapolations should be decreased by least 0.05 to
correct this bias. The precise correction will depend on how and where
the power law fit was made to the spectra, since there are many strong
blended emission lines at $>1250$\AA .

A second reason why the extrapolations might be biased is that shape
of the typical QSO spectrum in the \lyaf\ is much more complex than a
pair of power laws.  Figure 5 of \citet{press93}, Figure 6 of
\citet{vandenberk01}, Figures 4 \& 9 of \citet{telfer02}, Figure 3 of
\citet{bernardi03} and Figures 2 \& 3 of \citet{suzuki04a} all clearly
show that the region from \lyb -O~VI to \lya\ is dominated by the
wings of those two emission lines, and by lines near 1073~\AA\
(possibly blend of N~II, He~II and Fe~II, according to
\citealt{telfer02}, or Ar~I according to \citealt{zheng97}) and
1123~\AA\ (Fe~III). In some spectra 1073 and 1123 are weak, but in the
mean spectrum they are strong enough that all wavelengths are
influenced by one or more of these four lines, separated by three flux
minima, near 1050, 1100 and 1150~\AA .  The flux minimum near
1050~\AA\ between \lyb\ and 1073~\AA\ is especially hard to recognize
in spectra with a lot of absorption or low S/N, and although we were
looking for it, in some cases we missed it, as in the left panel of
Figure \ref{fxx}.

There are also weaker emission lines near 1176~\AA\ (C~III*,
\citealt{vandenberk01,telfer02,suzuki04a}), and possibly 1195 (Si~II,
\citealt{telfer02}) and 1206~\AA\ (Si~III, our HIRES spectrum) all in
the wings of the \lya\ line, and outside our DA region.

The sign of the bias in the continuum that comes from ignoring these
lines will depend on the details of the continuum extrapolation or
fit, and in the case of a fit, on whether the person making the local
continuum fit was aware of these lines, attempted to fit them, and had
spectra of wavelengths $>1216$~\AA\ where other emission lines suggest
the likely strengths of the \lyaf\ lines.

\citet{bernardi03} were the first to call attention to the emission
lines near 1073 and 1123~\AA , and they fit QSO continua with power
law plus three Gaussian functions, one each for these lines and \lya .

In \citet{suzuki04a} we used principal component analysis to predict
the shape of the \lyaf\ continuum and emission lines in individual
spectra, using the shape of the spectrum at wavelengths 1216 --
1600~\AA . The results were sometimes excellent, but other times poor,
in part because of the sensitivity to the flux calibration. The
predicted flux in the \lyaf\ of a QSO had an average absolute error of
9\%, with a range from 3 -- 39\%, which is too large an error for a DA
measurement, especially since we do not know whether the mean is
systematically too high or too low.

Amongst the methods that use the local flux in the \lyaf\ we note
\citet{fang98} who used the mode of the distribution of flux to
estimate the continuum level on our HIRES spectrum of one bright
QSO. This works best when the S/N is very high and the spectral
resolution is high enough to show regions with minimal absorption,
neither of which is the case for our Kast spectra.

\subsection{The Local Continuum Fits that we made}

The method that we use to fit continua is that normally used on high
resolution spectra with high S/N: fitting a different smooth curve to
each \lyaf\ spectrum.  This method is routinely used for measurements
of absorption lines, and for DA when the spectra have high resolution,
high S/N, and the \zabs\ is low enough that there are some regions
that appear to be absorption free.

\citet{rauch97} fit local continua to Keck HIRES spectra of 7 QSOs at
\zem = 2.5 -- 4.6 that differed widely in S/N.  They used spline fit
continua with rejection of 3$\sigma$ depressions.  They corrected
their continua upwards because some regions lack pixels without
absorption. The corrections came from the highest flux in each
simulated spectrum, from a box $10 h^{-1}$ comoving Mpc long for a
$\Lambda $CDM model. At $z = 2$ this correction was small, from DA
$=0.148$ to 0.154. We shall also use artificial spectra make
corrections to the whole of each QSO spectrum.

\citet{mcdonald00a} fit local continua to the same spectra used by
\citet{rauch97} plus one more.  They used IRAF to fit Spline3 or
Chebyshev polynomials to the continua, which were cut into 2 -- 4
pieces prior to the fits. They fit to the flux in portions of the
spectra that seemed free of absorption, with various orders of
polynomial. The fits were complicated because the flux calibration was
not good.  \citet{schaye03a} also fit local continua to 22 UVES and
HIRES spectra.

We know from our work on D/H measurement that we can fit the \lyaf\
continuum in high S/N HIRES spectra with an error of around 2\%
\citep{kirkman03}.  We fit a smooth curve by eye, using a b-spline as
a convenient way to store the result.  This motivated us to try to use
and calibrate the same method on the Kast spectra. We find that the
method works because the lower spectral resolution is compensated by
the much lower density of \lya \ lines at $z \simeq 1.9$, and because
we have a large wavelength range, extending to near C~IV, in each
spectrum.

We used a b-spline fit to each spectrum that started with one control
point every 50~\AA .  We added points, especially in the emission
lines, and we manually adjusted all points over the entire spectrum to
obtain our best guess at the unabsorbed continuum.  The continuum is
strongly influenced by the observed flux levels throughout a spectrum,
and by our perception of the strengths and shapes of all the emission
lines throughout the spectrum.  We did not explicitly assume that any
particular part of a spectrum was absorption free.

We paid close attention to the emission lines, especially those at
1073 and 1123~\AA\ \citep{suzuki04a}.  We sorted the spectra according
to the strength of these lines, and we attempted to produce a
consistent set of lines across the whole of each spectrum, out to
C~IV. We found that strong lines such as Si~II 1263, O~I/Si~II 1306
and C~II 1335 often indicated strong lines near 1073 and 1123.

Two of us reviewed the entire set of fits to the Kast and artificial
spectra all in one sitting, to try to make a consistent set of fits
for the entire sample. We repeated this exercise after we had made the
corrections following the first review.

\subsection{Errors in our Continuum Fits}

We now examine the errors in the continua that we fit to the
artificial spectra. In each case we know the true continuum level.  We
will make the important assumption that the continua on the Kast
spectra have similar errors.

To measure the error in our continuum fits, we define 
\begin{equation}
\rm F1/TC = (fitted~ continuum)/(true~ continuum).
\end{equation}
We measured F1/TC for each pixel in each artificial spectrum.  The
mean of the absolute fractional error in our continuum fits in the DA
region is \fmce \%. At a random wavelength in a random QSO, 16\% of
the pixels have F1/TC $>$ \ftch, and 16\% have F1/TC $<$ \ftcl.  The
standard deviation of F1/TC is about \ftcs \% in the DA wavelength
range.  At 1216 -- 1500~\AA\ the F1/TC standard deviation is \ftcms
\%.

\begin{figure}
\epsscale{1.0}
\plottwo{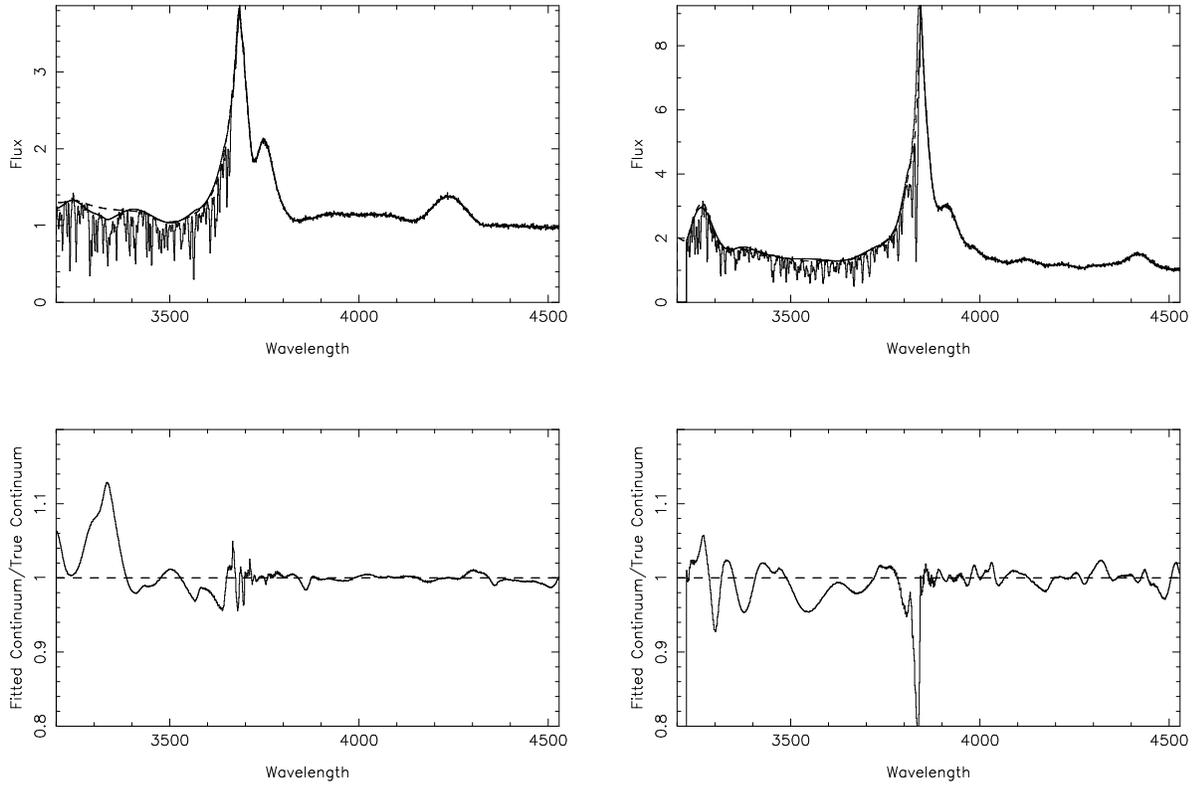}{d20d_q68.ps}
\caption{\label{fxx}
Examples of the continua that we fit to two artificial QSO spectra.
In the top panel for each QSO the solid curve is the true continuum
and the dashed line our fit.  Under each spectrum we show the fitted
continuum divided by the true continuum: F1/TC.  The spectrum on the
left has the largest error in the \lyaf\ of any spectrum with high
S/N.  The artificial spectrum on the right was selected at random, and
also happens to have high S/N.  }
\end{figure}

In Figure \ref{fxx} we show our F1 continuum fits to two artificial
spectra. For each we show both the true and the fitted F1 continuum
and the ratio F1/TC. These spectra illustrate several errors typical
of the continuum fits.  We often underestimate the amount of
absorption near the peak of the \lya\ line (right spectrum). We were
aware of this possibility but we still failed to anticipate the full
effect.  We also failed to give the emission lines enough
structure. For the left hand spectrum we failed to drop down between
the \lyb\ -- O~VI blend and the line at 1073~\AA\ that can be almost
as high as \lyb .

Although the S/N is higher in emission lines, the uncertainty in their
shape more than compensates. On average, the dispersion in F1/TC is
slightly larger in the emission lines.  In some QSOs the emission
lines have typical F1/TC (left spectrum in Fig. \ref{fxx}), but in others
the \lya\ line has the largest continuum error of anywhere in the
spectrum (right spectrum).  Errors are also larger at the UV end of
spectra where the S/N is lowest.

We classify our fits around the \lya\ emission line peak as
follows. For 34 QSOs the fits are good to excellent, and no worse than
elsewhere in the \lyaf . For 31 we fit too low, usually over a single
region 3-10~\AA\ (rest) wide (like the right spectrum in Fig. \ref{fxx}), but
sometimes over wider region.  For 3 QSOs we fit too high, and for 9 we
fit one region too high and another too low. We considered and
rejected using this knowledge to adjust our continua on both the
artificial and Kast spectra, because it would not change our results.

The errors in the continuum fits were correlated over a variety of
lengths, determined by the number of b-spline points that we chose to
use. We typically could justify using more points where the S/N was
high in the \lyaf , in the red, and where there are many pronounced
emission lines.  Although there was no particular scale of
correlation, we often saw strong correlations over 10--40~\AA\ in the
rest frame in the \lyaf\ region for most spectra and over a few \AA\
in the \lya\ line.

The continua on some of the artificial spectra are poor, with errors
of 10 -- 20\%, all across the \lyaf . Of the four with the largest
errors, two have low S/N, but two others have intermediate S/N $\simeq
5$.

\section{Correcting the Continuum Levels using the Artificial Spectra}
\label{se}

We measured DA in the DA wavelength range: 1070-1170~\AA .  We
examined plots of our Kast spectra, stacked in rest wavelength, to
help us make these choices.  We chose these wavelengths to avoid the
proximity effect, the wings of \lya\ where the continuum is changing
rapidly, the \lyb -- O~VI blend (1025.72, 1031.9, 1037.6~\AA ) and the
local flux minimum near 1050~\AA\ that is hard to recognize.
Associated absorption that was falling towards the QSO at
$-3000$~\kms\ (near the maximum velocity ever seen) would have its
O~VI 1037 at 1048.09~\AA .  In some spectra the continuum fits get
noticeably worse at $<1070$~\AA , and we will see that the standard
deviation increases at $<1070$~\AA .  For a \lya\ line, \lamr =
1170~\AA\ is 11300~\kms\ or 170 comoving Mpc from a QSOs at the mean
\zem , sufficiently far that we do not expect a significant influence
from the observed QSO.  Later in this section we will see that we
systematically fit the continuum too low in the range 1170 -- 1216~\AA
.

The mean of the DA values in all pixels in the artificial spectra,
averaged over the DA region wavelengths and all artificial spectra,
and using the true continua, is \dastruem. This value applies to a
mean \lya\ $z=1.924$.

\subsection{Correcting the continua of the Artificial Spectra using SNR2}

We defined a new variable, SNR2, to act a measure of the effect of
photon noise on the data quality. It is an indicator of the data
quality on large scales, and represents the smoothed S/N that we would
have measured in absence of absorption and emission lines. We ignore
both absorption and emission lines because to first order we do not
expect our continuum level estimates to be better or worse near a
narrow absorption line and we found that the continuum fits are
slightly worse in the emission lines.

We defined SNR2 as a smoothly varying function of wavelength, with
values similar to the S/N at those wavelengths. It is not smoothed
S/N, since this is depressed by absorption lines. Rather it is the
flux and S/N in the relatively unabsorbed regions of the spectrum that
most influence the continuum fitting accuracy. SNR2 is a linear fit to
the S/N values that we measured at the two reference wavelengths.

\begin{figure}
\epsscale{0.7}
\plotone{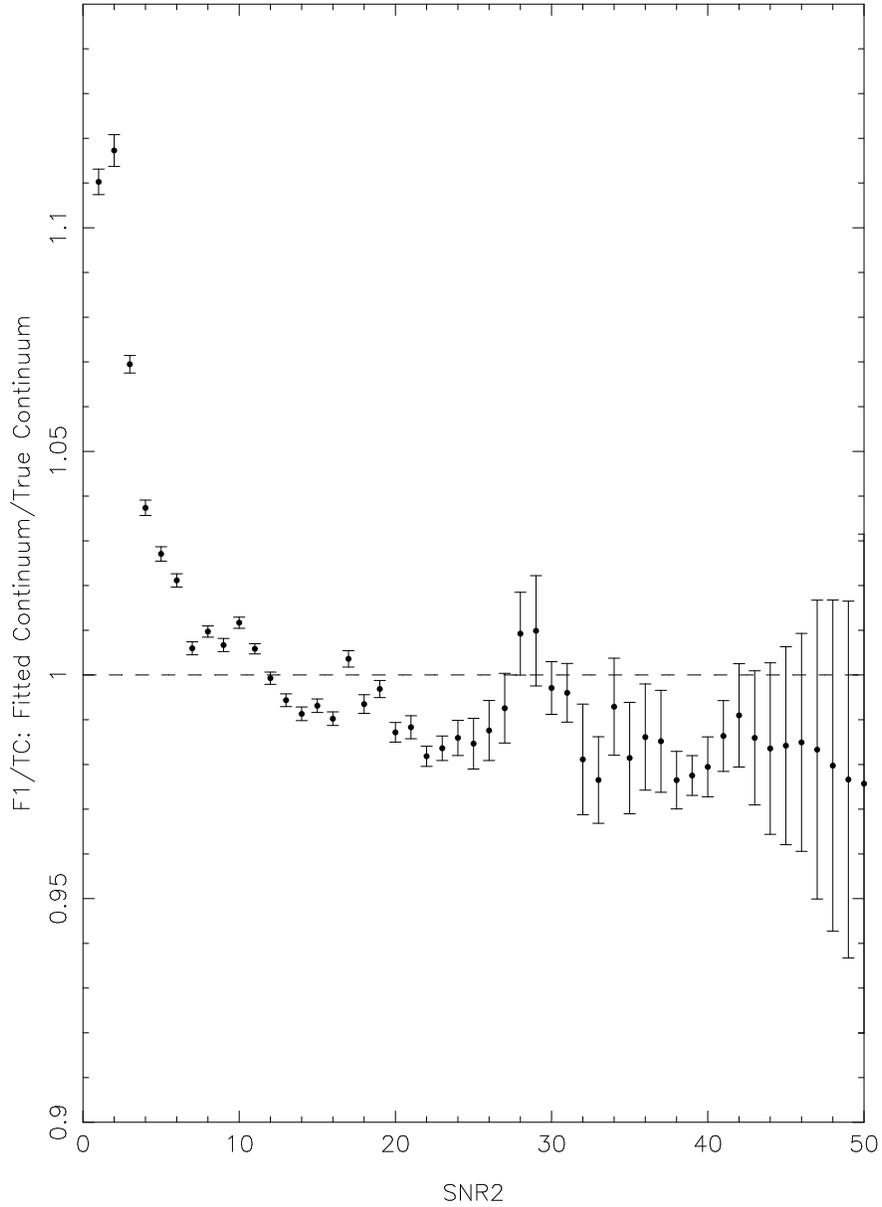}
\caption{\label{fxxx}
The fractional error in the continuum level as a function of the data
quality. The points show the ratio of our fitted continuum to the true
continuum, F1/TC, in the artificial QSOs as a function of SNR2.  The
error bars on each point indicate the error on the mean value of the
fitted/true continuum ratio.  The F1/TC values for each pixel have a
much larger dispersion, especially for small SNR2.}
\end{figure}

In Figure \ref{fxxx} we show F1/TC as a function of SNR2. We show the mean
F1/TC for all pixels (often over 1000) in a given SNR2 range, and we
use all pixels from 1050 -- 1070~\AA , which extends to 20~\AA\ lower
wavelengths than the DA region. The means show correlation between
adjacent bins, since the F1/TC in the spectra are also correlated over
tens of \AA , and hence over a range in SNR2, which varies smoothly
with wavelength. The plot shows that the F1 continua are
systematically too high at SNR2 $< 6$, and usually too low at SNR2
$>12$.  This type of error is not unexpected.

We made new continua, labelled F2, by dividing the F1 continua on both
the Kast and artificial spectra by the factors show in Figure \ref{fxxx}.

The correction worked as expected on the artificial spectra.  The DA
we measured using the initial continua F1 was \dasfonem, significantly
too large.  The DA measured using the F2 continua was \dasftwom, which
is 0.990 of the value for the input artificial spectra: \dastruem.
The DA4(F1) was too large by 5.2\%.  We checked that when we applied
the correction to 1050 -- 1170~\AA\ the total DA was 0.1583, which is
0.1\% less than the DA $=0.1585$ for this wavelength range in the
input artificial spectra.

The effect is similar for the Kast spectra: We had DA = \darfonem\
using the F1 continua, and we find DA = \darftwom\ using the F2
continua. DA(F1) was too large by 6.8\%, similar to the excess for the
artificial spectra, but not identical because the distribution of flux
as a function of SNR2 is different.

\subsection{Correcting the continua of the Artificial Spectra using SDA}
We define a second variable which also indicates regions of the
spectra where we might have made systematic errors in the continuum
fits.  SDA is smoothed DA, obtained by smoothing the flux with an
exponential filter with FWHM 25~\AA\ rest. This length is similar to
the scales on which we see strong correlations in the continuum
errors.

\begin{figure}
\epsscale{1.0}
\plotone{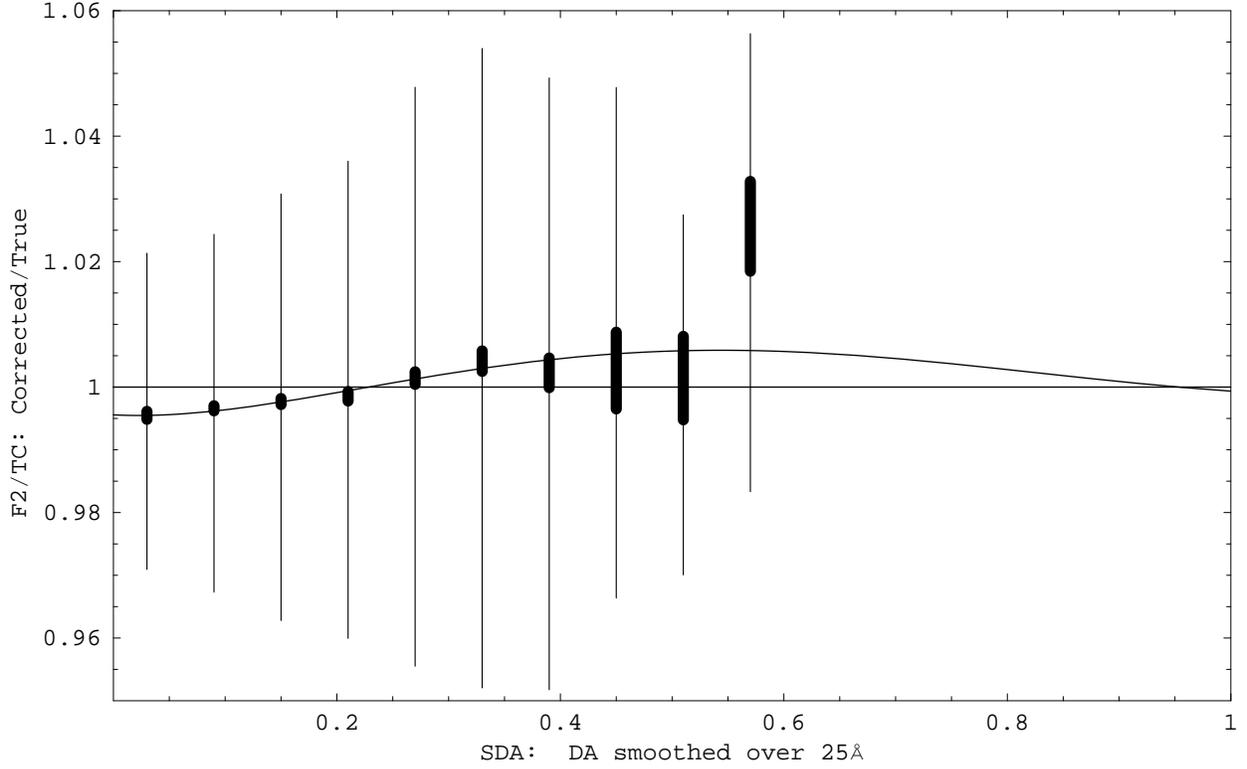}
\caption{\label{fl}
Fractional error in the partly corrected continuum level as a function
of the mean amount of absorption.  The points show the ratio of the
continuum corrected for the SNR2 correlation to the true continuum in
the artificial QSO spectra, F2/TC, as a function of SDA.  The thick
bars indicate the mean and the error on the mean at each SDA value.
The thin bars show $\pm 1\sigma $ for the F2/TC evaluated in all
pixels in the SDA interval. The smooth curve is the function we used
to implement the SDA correction.}
\end{figure}

In Figure \ref{fl} we show the F2/TC as a function of SDA. We see a
systematic trend that indicates that F2 is too high by about 1\% on
average in regions of the artificial spectra that have $0.3 < $ SDA
$<0.6$. We used the smooth curve to approximate the corrections that
we made to F2 to remove this trend. We show how we extrapolated the
curve to higher SDA values that occur in some parts of the Kast
spectra, but not in the artificial spectra, because the artificial
spectra differ from the Kast spectra.

We label F3 the continua that we corrected for the SDA correlation (F2
to F3) after we had corrected them for the SNR2 correlation (F1 to
F2).  For the artificial spectra, the mean DA following the correction
to F2 was \dasftwom\ and after the correction using SDA the F3
continua gave DA \dasfthreem.  The change is 1.1\%, and leaves the DA
nearly identical to the known value, \dastruem, as required by the
definition of the corrections.

For the Kast spectra, the change is a 1.2\% increase, from DA(F2) =
\darftwom\ to DA(F3) = \darfthreem.  This is slightly different from
the change to the artificial spectra because the distribution of the
flux as a function of SDA can differ from that for the artificial
spectra.

In Figure \ref{flv} we see that F3/TC is a well behaved function of rest
wavelength.  The thick bars show the mean values, from all QSOs
(usually 77) that contribute at that wavelength.  The mean F3/TC
across the DA region is near 1.0 by definition of the corrections.
About 16\% of pixels have F3/TC $<0.96$ and 16\% are $>1.03$, with
little variation across the DA region.  We are not surprised that the
mean F3/TC values are correlated over many pixels and regions of
around 60~\AA\ because the continuum errors F1/TC were also correlated
over such large scales.  We also see a tendency for the F3/TC to be
too low in the interval 1170 -- 1216~\AA . We saw that we fit the
continuum too low in this region for some QSOs.

In Figure \ref{fdli} we show the standard deviation of the F3/TC values,
$\sigma $(F3/TC), as a function of \lamr .  The value is nearly
constant, around 1.3\% at $>1220$~\AA . It peaks near the peak of \lya
, at 4.5\%, declining slowly as we move down the blue wing of \lya
. It rises again below 1070~\AA , reaching a level of 5 -- 6\%.  We
used this Figure to help us choose the DA region.

\subsection{Effects that may remain uncorrected}

Here we discuss two of several possible sources of error in our
continuum estimates that the SNR2 and SDA corrections may have missed.

First, there are errors in the flux values in our QSO spectra, due to
the limitations of the flux calibration procedure applied to the
observed spectra.  There are numerous sources for such errors, which
we describe in detail in \citep{suzuki03b}.  Fortunately, many flux
calibration errors in low resolution spectra vary smoothly over large
scales $>50$~\AA, and they will mostly have been absorbed as
adjustments to the continuum fit, reducing their effect on the
DA. Some of these errors might vary randomly in sign from QSO to QSO,
and with wavelength, leading to a minimal net effect in our sample,
but others are systematic.

An example of a systematic flux error is the strong atmospheric ozone
absorption that varies in strength with a period of about 25~\AA , and
increases a lot below about 3310~\AA\ \citep{schachter91}. We did not
explicitly remove the ozone absorption from either the standard stars
or the QSO spectra, hence we expect to see this pattern in the \lyaf\
absorption.  In Figure \ref{flvii} we show DA as a function of
observed wavelength.  Some variations can be seen at rest wavelengths
$<$ 3250 \AA, which are probably due to ozone absorption.  However, we
note that the variations go both high and low, because we generally
fit continua through ozone absorption, not above or below it, so the
net effect on our estimated DA should be minimal.

Second, it is possible that the artificial spectra differ from the
Kast spectra in some way that is difficult to notice, but nonetheless
very important.  We know that the artificial spectra lack strong \lya\
lines from high density regions that have high \nhi .  In the Kast
spectra these lines alone have DA $=1$\%.  What if the Kast spectra
also have some smoothly varying absorption that covers many
wavelengths, but is not included in the simulations?  Since we
examined the whole of each spectrum, from near C~IV to below \lyb , we
hope that we have correctly accounted for such hypothetical
absorption.  However, the F3/TC will not tell us if we erred.

In summary, we are reasonably confident that we have accounted for the
main systematic effects.
\section{DA in Kast spectra}
\label{sf}

We applied exactly the same corrections to the \lyaf\ of the Kast
continua as we applied to the artificial ones to make the equivalent
of F3 for the Kast spectra.  We will no longer mention the uncorrected
continua on the Kast spectra.  Hence, the DA in a pixel, which we will
label DA0, is the flux divided by the corrected continuum fit, F3.  Of
course we can not show you ratios like F3/TC for the Kast spectra
because we do not know the true continuum levels.

In Figure \ref{flix}
we show two statistics from the DA0 values.  The thin vertical lines
show the $\pm 1 \sigma $ of the DA0 values.  The center of the short
solid bars show the mean DA0 from all DA0 in the 4.5~\AA\ rest frame
wavelength regions, which we call DA1.  The length of the thick bars
show the $\pm 1\sigma $ errors on those means, calculated assuming
that the DA0 values were random normal deviates: $\sigma (\mu ) =
\sigma (DA0)/\sqrt{n}$, where the number of pixels per bin is $n
\simeq 900$. We know that this underestimates the error on the mean,
both because there are about 3 pixels per resolution element, and the
DA0 values are far from normal deviates.  We also show the DA
wavelength range.  We make several points from this plot that guide
our decisions on how we will measure the overall mean DA4 value.

First, at 1225--1500~\AA\ there is clearly absorption due to metal
lines.  The mean is around 2\%, systematically increasing to smaller
wavelengths.  We will discuss this below.

Second, the DA1 values rises smoothly as we cross the peak of the
\lya\ line.  The transition across the \lya\ emission lines is wide,
from about 1210 -- 1225~\AA , a range of around 3700~\kms . We see
that the DA1 near the wavelength of the peak of \lya\ (the lower end
of the arrow labelled \lya\ emission) is near the mean of the DA1 on
either side of \lya .  This trend comes from the proximity effect,
modified by the errors in the \zem\ values that we used and the large
errors in continua near the peaks of the emission line.  We would need
to improve these two factors before we could use these data to measure
the proximity effect.

Third, the DA1 is approximately constant across the DA region.

\begin{figure}
\epsscale{1.0}
\plotone{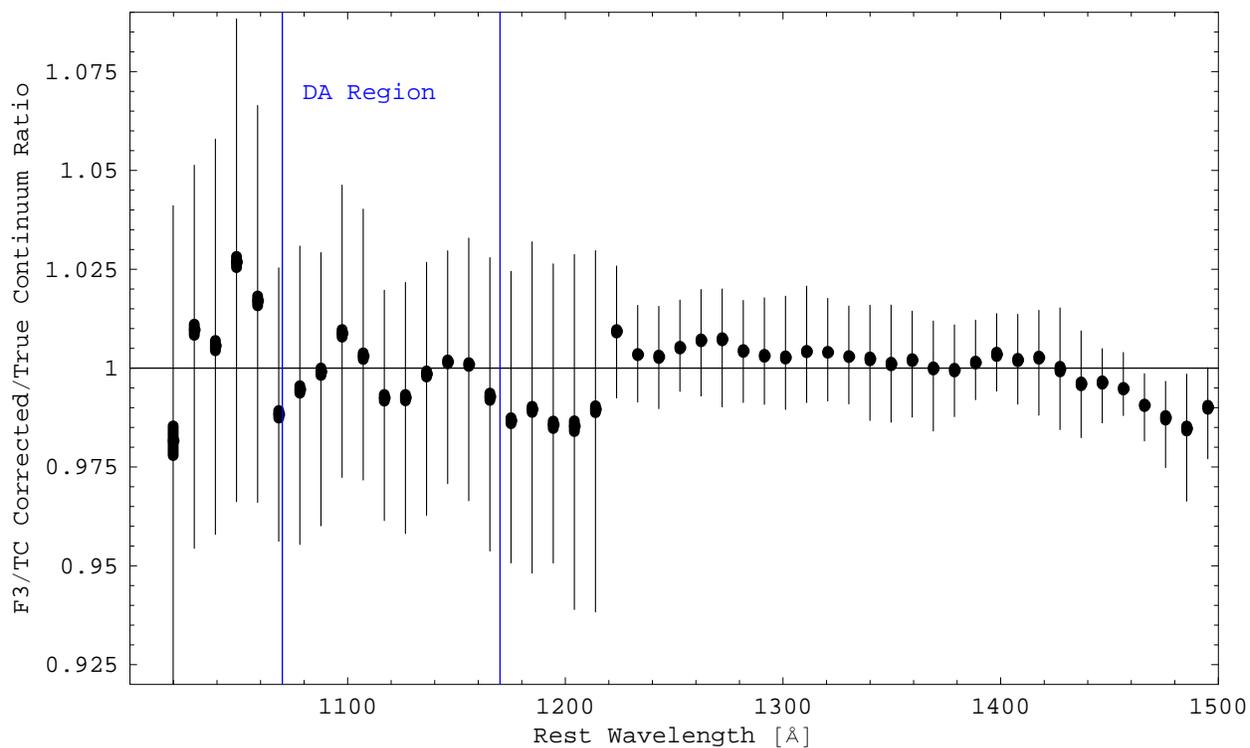}
\caption{\label{flv}
Fractional error in the fully corrected continuum level as a function
of rest wavelength.  We show the ratio of our F3 continuum to the true
continuum in our sample of artificial spectra.  The F3 continuum has
our SNR2 and SDA corrections applied for $\lambda_r < 1216$ \AA, while
for higher wavelengths the F3 continuum is the original F1
continuum. The two vertical lines show 1070 and 1170~\AA , the
boundaries of the region where we measure DA.}
\end{figure}

\begin{figure}
\epsscale{1.0}
\plotone{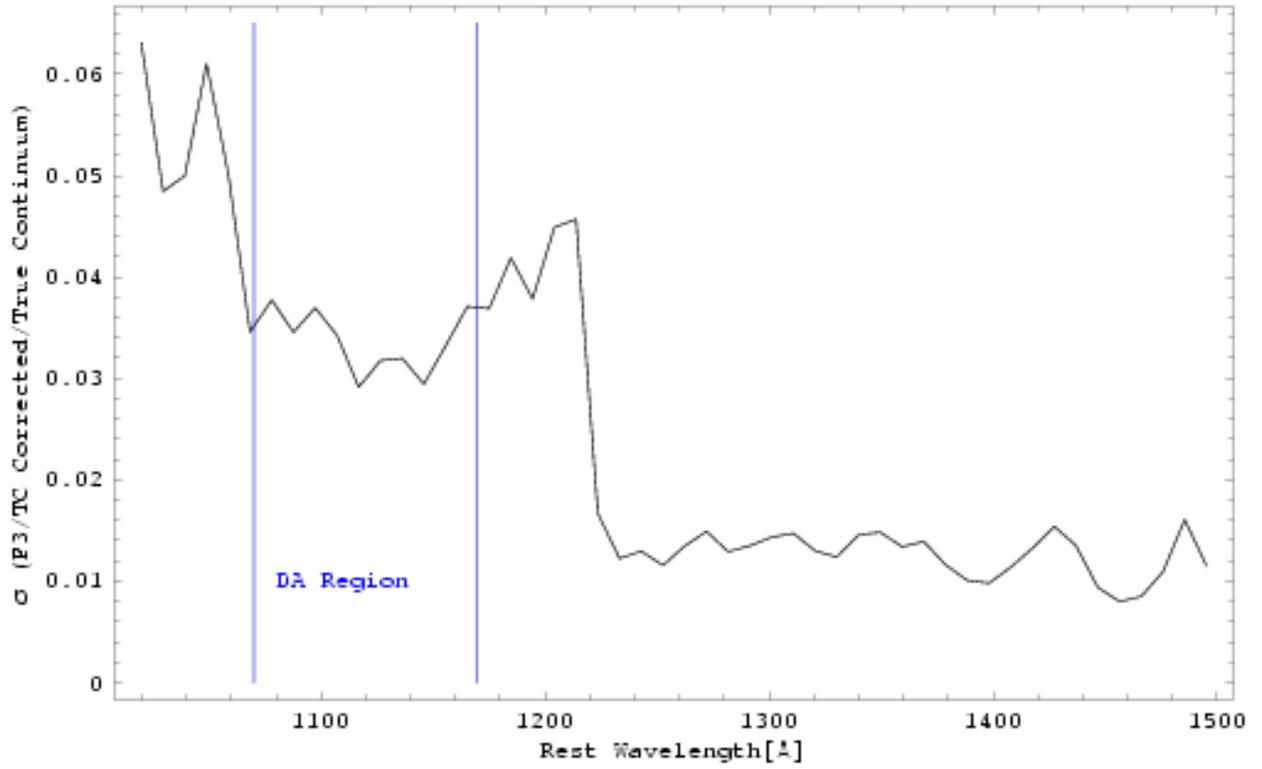}
\caption{\label{fdli}
The standard deviation of the fractional error in our continuum fits
to the artificial spectra, F3/TC.  Our continuum fits appear to be
well behaved in the region we use to measure DA: $1070 < \lambda_r <
1170$ \AA.}
\end{figure}

\begin{figure}
\epsscale{1.0}
\plotone{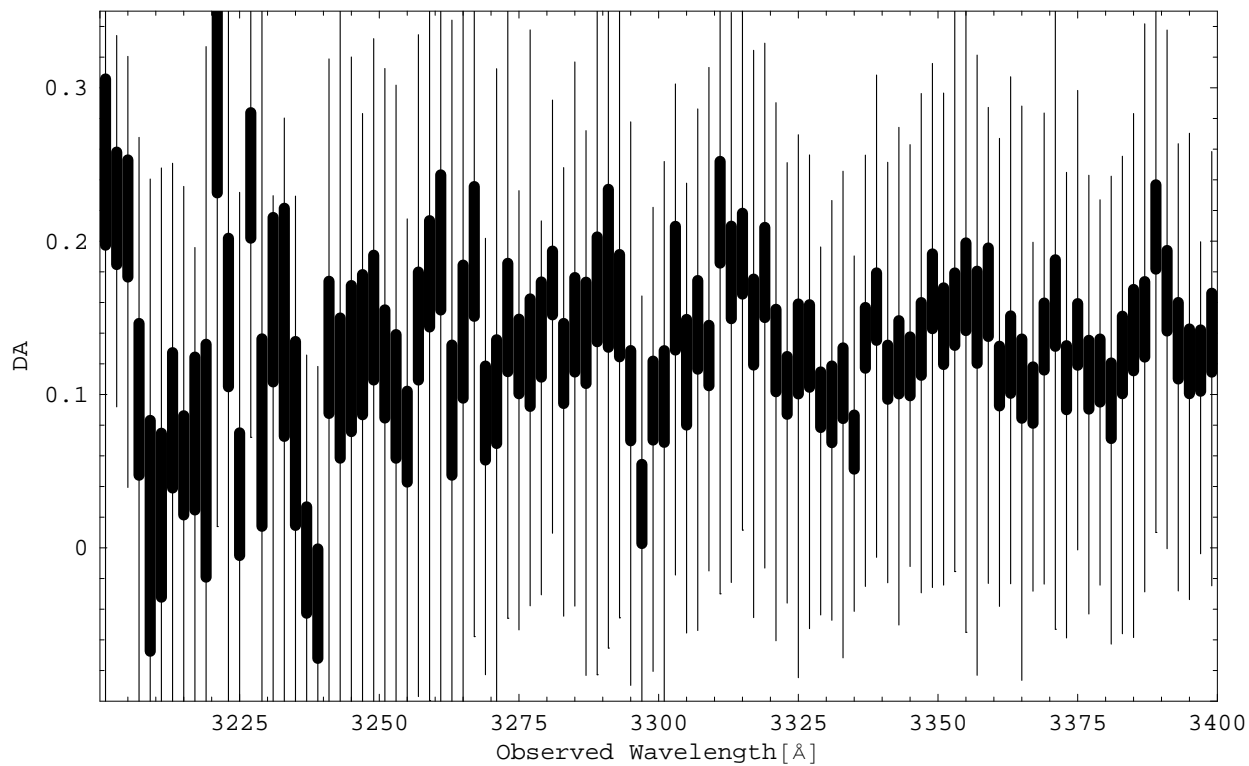}
\caption{\label{flvii}
The total amount of absorption, shown as DA, in the Kast spectra as a
function of observed wavelength where Ozone absorption is expected.
We expect the most Ozone absorption, and the highest DA values, near
the marked wavelengths, 3198, 3220, 3252, 3280 and 3310~\AA .  As in
previous Figures, the center of each thick bar is at the mean DA value
in the bin, and its length is the $\pm 1\sigma $ error on the mean,
while the thin lines the $\pm 1\sigma $ of the DA values per pixel.  }
\end{figure}

\begin{figure}
\epsscale{1.0}
\plotone{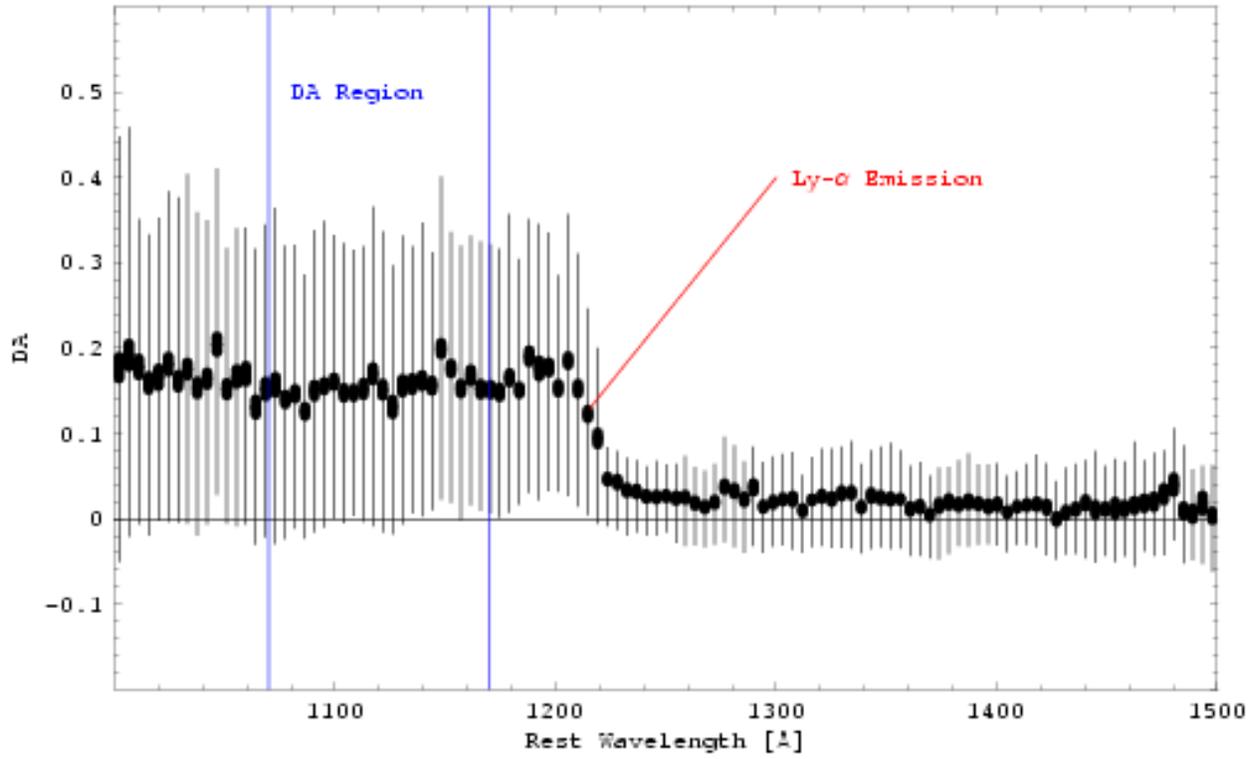}
\caption{\label{flix}
Total absorption in the Kast spectra, shown as DA, as a function of
rest wavelength.  The DA includes absorption by \lya\ in the IGM, in
LLS, DLAs and metal lines.  The thin lines show the $\pm 1\sigma $
values for the DA0 per pixel. The heavy lines show the mean and error
on the mean DA in the 4.5~\AA\ bins, the DA1 values.}
\end{figure}

Fourth, the DA1 is higher in 1190 -- 1205~\AA . This might be related
to the continuum fitting errors which made the F3/TC too low for 1170
-- 1220~\AA\ for the artificial spectra (Figure \ref{flv}).

Fifth, the DA1 is higher at 1000 -- 1045~\AA . Some of this might be
from continuum errors that increase as the S/N decreases.  Figure \ref{flv}
showed that the F3/TC values had a large $\sigma $ at these
wavelengths, and four of the five mean F3/TC values were slightly too
high.  Rather the DA1 is probably higher because of absorption by
\lyb\ and perhaps some O~VI.

Lastly, there is some extra remaining dispersion in the DA1 in the
\lyaf\ because we have not yet removed the tendency for DA to increase
with $z$.

\subsection{How DA Changes with $z$}
In Figure \ref{flx} we show DA as a function of the \lya\ absorption
redshift.  We give the values in Table \ref{tabd}.  The thin lines are
again the $\pm 1\sigma $ of all the DA0 (pixel) values in each $z$
bin, while the thick bars show the mean DA values, $\rm \mu (DA)$ and
their errors $\sigma (\mu )$, which we have underestimated.  In Table
\ref{tabd} we also list the usual $1 \sigma $ confidence interval for
the DA values per pixel. We give the critical DA values, where 15.8\%
of the values are below the lower value (column 4) and 15.8\% are
above the larger value (column 5).  We see a slight increase in the DA
with increasing $z$, as expected from well known counts of the number
of lines per unit redshift, a trend first found by \citet{peterson78}.
We fit this trend with
\begin{equation}
\label{daz}
DA(z) = A \left((1+z)/(1+1.9)\right) ^{\gamma },
\end{equation}
with $A= 0.147$, and $\gamma=$\zfitgamma.  These values are slightly
different from those for the input artificial spectra: $A=0.150$ and
$\gamma = 2.07$.  We do not quote errors on these parameters because
the range of $z$ covered is very small and the slope $\gamma $ is not
well determined.  We will concentrate instead on estimating the DA at
the mean $z$ for the sample.

\section{\lya\ Absorption from High Column Density Lines}

Some of the absorption in the DA region is from the \lya\ lines of
systems with high H~I column density, especially including LLS and
DLAs.  We now estimates how much, because these absorbers are much
harder to include accurately in the numerical simulations that have
the large boxes needed for the IGM. We expect that this is a temporary
situation, since we would prefer to use simulations that include all
the main features of the IGM and galaxies that are responsible for the
absorption in QSO spectra. They should include realistic LLS and DLAs,
with realistic velocity structure, temperatures and metal abundances,
giving realistic metal lines.

We introduce 
\begin{equation}
DA5(z) = N(z) W_r(1+z)/\lambda_r 
\end{equation}
as a general estimator of the amount of absorption from lines with a
rest frame wavelength \lamr , a density of $N(z)$ lines per unit $z$
and a mean rest frame equivalent width $W_r$~(\AA ). The $(1+z)$
factor converts $W_r$ to the observed frame, and for \lya\ there are
1215.67~\AA\ in the observed frame per unit $z$.

LLS with \lnhi $>17.2$~\cmm\ have Lyman continuum optical depth $>1$
\citep{tytler82} and a density $N(1.9) = 1.4 \pm 0.5$ from Figure 2 of
\citet{stenglerlarrea95}.  The fractional error is huge because HST
has obtained spectra of few QSOs that could show Lyman limits around
2650~\AA .  We measure a mean $W_r = 3.0 \pm 0.5$~\AA\ for 66 LLS and
DLAs measured in Kast spectra, and calibrated with 13 systems that we
also observed with HIRES \citep{burles97}.  These LLS and DLAs were
detected as Lyman limits at $2.4 < z < 4.1$.  We ignore possible
evolution in this mean $W_r$.  We obtained the error by summing three
terms in quadrature: 0.2~\AA\ from the calibration, 0.3~\AA\ from the
sample size and 0.3~\AA\ for possible bias in the sample.  The $W_r$
values have an approximately exponential distribution for small $W_r$
values, with an excess at the largest values from DLAs
\citep{sargent80}.

Using the values above, we find the contribution to the DA from \lya\
lines at $z=1.9$ in systems with \lnhi $>17.2$~\cmm\ is
\begin{equation}
DA6s = 1.0 \pm 0.4\%, 
\end{equation}
where nearly all of the error is from the uncertain number of LLS at
$z=1.9$.  Here, and elsewhere, the suffix ``s" refers to a value for
$z=1.900$.  Too few LLS are known at redshifts 1.5 -- 2.5 to determine
how the number per unit redshift changes with redshift.

We checked this result using only the DLAs with \lnhi $>20.2$~\cmm ,
 or $Wr > 10.3$~\AA .  The DLAs have N(z=1.9) $=0.20 \pm 0.04$ from
 \citet[Fig. 11]{storrie00}, and their mean $W_r = 17.78$~\AA , from
 their Eqn. (3) and \citet[Eqn. 3]{wolfe86}.  This mean is for DLAs
 observed at $1.5 < z < 4$.  They have DA $=0.85 \pm 0.17$\%, where we
 have ignored the error on the mean $W_r$. We expect this DA to be
 smaller than that for all LLS, but the difference is less than we
 expected.  Perhaps the DA6s value is too small. The LLS and
 DLAs have independent normalization, each has a large statistical
 error, and we have ignored evolution of the $W_r$ values.

\section{Absorption from Metal Lines}
\label{sfm}

To measure absorption by metal lines alone we now introduce 
\begin{equation}
DM = 1-\left< F\right>,
\end{equation}
where $F=1$ if there is no metal line absorption.  Since we defined DA
to included all types of absorption, DM $\leq $ DA in the DA
region. We measured DM in the \lamr\ range 1225 -- 1500~\AA , and we
extrapolated to estimate a value for DM in the DA range 1070 --
1170~\AA .

In Figure \ref{fcx} we show the DM as a function of rest wavelength.
We measure DM using the original continuum fits to the Kast spectra,
without the corrections for correlations with SNR2 and SDA that we
determined for the \lyaf .  Before making this plot we have measured
the DM at the same wavelengths in the artificial spectra. Except for
continuum level errors, this should be identically zero, since there
are no metal lines in the artificial spectra. We saw slight absorption
at all wavelengths, showing that we typically place the continuum too
high by 0.5\% (about 0.15$\sigma $ per pixel), and we subtracted this
from the Kast spectra to give the DM that we show.  A straight line
fit to these DM values from $1225 < $ \lamr\ $ < 1500$~\AA\ gave:
\begin{equation}
DM1(\lambda _ r) = 1.585 - 2.67135\times 10^{-4}  (\lambda _r -1360)\%,
\end{equation}
where the \lamr\ value is in \AA , and DM1 increase slightly
as \lamr\ drops.  However, a fit as a function of observed wavelength,
\lamo\ (\AA ) gives
\begin{equation}
DM2(\lambda _ 0) = 1.576 + 9.661 \times 10^{-4} (\lambda _o -4158)\%,
\end{equation}
which has a very similar mean near the wavelength, but now DM2
decreases as \lamo\ decreases.  For the Kast spectra in the range 1225
-- 1500~\AA\ the mean \lamo = 4158~\AA .  We list this and other
measurements of the metal absorption in Table \ref{tabc}.

We expect the DM1(\lamr\ ) and DM2(\lamo\ ) increase with decreasing
$\lambda $ for two reasons: additional lines are included at smaller
\lamr , and the density of C~IV lines increases at smaller redshifts,
or observed wavelengths \lamo .  However, the difference in the slopes
of DM1 and DM2 indicate that the trend is not strong.  Plots of these
trends are not too helpful because they are dominated by the huge
dispersion in the DM values, which makes it harder to measure the
slope of DM with wavelength.

We define DM3($\lambda )$ as the mean DM in a segment of the spectrum
of a QSO that is 121.567~\AA\ wide in the observed frame,
corresponding to $\Delta z = 0.1$ for \lya .  The DM3 values
illustrate the huge dispersion in the DM across large parts of a
spectrum.  We began the first segment for a QSO at 1225~\AA\ in the
rest frame, the next one started where the first one ended, and the
last one ended before the minimum of the maximum observed wavelength
and 1500~\AA .  We ignored the remaining part of each spectrum that
gave incomplete segments covering $<121.567$~\AA\ near the maximum
wavelength.

In Figure \ref{dm2z} we show the DM3 values from the Kast spectra as a
function of observed wavelength, expressed as redshift for \lya , to
aid comparison with Figure \ref{flx}.  The mean of the DM3 values is $1.87
\pm 0.13$~\%, and the $\sigma (DM3) = 2.5$\% excluding photon noise,
or 2.6\% with the photon noise.  The distribution of DM3 values is
decidedly skew, with a long tail to huge DM3 values. Correlations
amongst the DM3 extend right across a spectrum, because absorption
systems with high \nhi\ values have many lines, these lines are
strong, and they occur all over one spectrum.  Moreover, systems are
strongly clustered on scales up to 600~\kms\ \citep{sargent88a}.

\subsection{Measurement of DM using Line Lists in Published Spectra}

We have measured DM4 values using the lists of absorption lines
published by \citet{sargent88a} for 26 QSOs with $1.7 <$ \zem\ $<
2.3$, excluding Q1510+115 which is BAL.  Eleven of these QSOs are also
in our Kast sample.  We defined
\begin{equation}
DM4 (\lambda ) = \sum _{\lambda i} ^{\lambda i +121.567}
                 W_{obs}/121.567~\rm \AA ,
\end{equation}
as an estimator of mean DM from all metal lines in bins of width
121.567~\AA\ in the observed frame.  As with the DM3 from the Kast
spectra, the first bin started at the maximum of the minimum observed
wavelength \citep[Table 1]{sargent88a} and 1225~\AA\ in the rest
frame, and the last bin ended prior to the minimum of the maximum
observed wavelength and 1500~\AA .  Each QSO contributed 4 -- 7
segments, and we ignored the partial segments.  The mean wavelengths
were \lamo $=4125$~\AA\ and \lamr $=1358$~\AA .  We took the observed
frame equivalent width values $W_{obs}$ from their Table 3. The
results, in our Table \ref{tabc}, are similar to those from the Kast spectra:
the mean of the 153 DM4 values was $1.67 \pm 0.22 $~\% and the
$\sigma (DM4) = 2.74$\%.  The mean is nearly identical to that for DM1
from Kast.

DM3 and DM4 differ in several ways.  The most obvious difference
between the distributions of DM3 and DM4 is that DM4 has a larger
fraction of segments with DM $< 0.5$\%, those with few or no
absorption lines. The DM3 values are effected by continuum errors,
photon noise and weak absorption lines, all three of which are less
prominent in the DM4 values. Typical weak lines in \citet{sargent88a}
have $W_{obs} > 0.25$\AA\ that individually give DM $= 0.2$\%.  Lines
that are weaker than this will have been missed from DM4.  The mean
values are similar because weak lines do not produce a large part of
the total absorption.  Given this, the DM4 might have comparable
accuracy to the DM3, since the smaller sample for the DM4 will be
partly compensated by the lower sensitivity to continuum errors and
photon noise.

We have fit the DM4 values as a function of both \lamr\ and \lamo\ giving:
\begin{equation}
DM5(\lambda _ r) = 1.662 - 5.511 \times 10^{-3}  (\lambda _r -1360)\%,
\end{equation}
and
\begin{equation}
DM6(\lambda _ 0) = 1.649 - 7.136 \times 10^{-5} (\lambda _o -4158)\%.
\end{equation}
As with the DM1 and DM2 fits to the Kast metal lines, the mean values are
similar and the slopes differ. This time the fit to the \lamr\ has the
shallower slope.

\subsection{Absorption by C~IV}

\citet{sargent88a} and \citet{steidel90a} found that the number of
C~IV absorption lines increases as $z$ decreases, and recent
measurements show the same trend \citep{misawa02}.

We have estimated the DM due to C~IV alone using Equation (12) of
\citet{sargent88a}.  By definition, the total absorption from the
stronger of the C~IV doublet lines, 1548, is $N_*W_*$, per unit
$z$. We obtained this by integrating the $n(W_*)$ equivalent width
distribution over all $W_r$, from zero to infinity. This involves an
extrapolation to $W_r < 0.15$\AA\ that is probably not much in error
because the strong lines that were measured by \citet{sargent88a}
dominate the total absorption.  Then
\begin{equation}
DM7(1548, z=1.957) = N_*W_*(1+z)/1548 = 0.404 \pm 0.074 \%,
\end{equation}
where the $(1+z)$ factor converts $W_r$ to an observed equivalent
width, and $N_*$ is the number of lines per unit of $z$ for this line,
1548~\AA\ in the observed frame. We use $W_* = 0.46\pm 0.04$\AA\ and
$N_* = 4.60 \pm 0.74$ for their sample A4, and we assume that the
sample is at about the redshift of their sample S2, 1.957.  From their
Figure 10, at this $z$ the mean $W_r(1548)/W_r(1550) = 1.42 \pm 0.10$
hence,
\begin{equation}
DM7(C~IV, z=1.957) = DM(1548)+DM(1550) = 1.704W_r(1548) = 0.69 \pm 0.15\%.
\end{equation}

The absorption due to C~IV increases slightly as wavelength decreases.
We make two simple assumptions to measure this trend.  First, we scale
the DM7(C~IV) by a factor $[(1+z)/2.957]^{\gamma }$ where
\citet{misawa02} find $\gamma = -0.58 \pm 0.46$ for their sample EM15
that includes 136 C~IV systems with $W_r(1548) > 0.15$~\AA .  Second,
we convert the \zabs\ for C~IV(1548) to \lamr\ for \lya\ by assuming
that all our QSOs are at the mean \zem $=2.17$: $(1+z_{1548})1548 =
\lambda _r (1+z_{em})$.  We find that $DM(C~IV) = 0.71$\% at \lamr =
1360~\AA , which is 0.45 of the total DM1(1360).

\subsection{Extrapolating DM into the DA region}

At the mean \lamr = $1120$~\AA\ where we measure DA, the four
extrapolated DM values, DM1, DM2, DM5 and DM6, have a mean of
\begin{equation}
DM1-6(1120) = 1.92 \pm 0.42\% 
\end{equation}
where the error is $\sigma /2$.

The DM increases in part because lines with smaller \lamr\ start to
contribute in the spectrum at $\lambda < \lambda _r $.  In the range
$1225 < $ \lamr\ $ < 1500$~\AA\ these include Si~IV 1394, C~II 1334,
O~I 1302, Si~II 1260 and N~V 1229.  In the wavelength range 1070 --
1170~\AA\ such lines include Si~II 1193 and especially Si~III 1206.

We increase our estimate of the DM in the \lyaf\ by a factor of $1.2
\pm 0.1$ to account for the increasing number of lines as a function
of decreasing \lamr\ because we suspect that these lines are not
adequately accounted for by the trends with \lamr .  The adjustment
has a large error because it is based on our examination of only 9
spectra of high resolution at these redshifts.

The DM we will use is then
\begin{equation}
DMs(\lambda _ r =1120) = 2.3 \pm 0.5\%,
\end{equation}

We estimated that had we been able to measure DMs values, in the \lyaf
, in bins of $\Delta z = 0.1$, those values would have $\sigma =
3.1$\%, which we obtain from $\sigma \propto (mean~DM)$, using $\sigma
(DM3)=2.5$\% for mean~DM3 $= 1.87$\%.

We find that the metals account for $15 \pm 4$\% of the total DA at
$z=1.9$. This is less than the estimate of \citet{rauch97}, from a few
QSOs, that about 22\% of the total DA is from metals at $z \simeq
2$. If \citet{rauch97} used a path length of about $\Delta z = 0.9$
from about 3 QSOs, then their DM would have a relative error of order
$3.1/(2.3 \times \sqrt{9}) = 0.45$\%, sufficient that the difference
from our estimate would not be significant.
\section{Dispersion of DA Values}
\label{sg}

We have long known that there is a lot of scatter in the amount of
\lyaf\ absorption present in different low \zem\ QSOs: \citet[\S
7]{carswell82}, \citet{tytler87a}, \citet[Fig. 15]{kim01}, and
\citet[Fig. 8]{kim02b}.  \citet{schaye03a} found that the $\chi ^2$
per degree of freedom of DA measured in intervals of about $\Delta z =
0.2$ was 5.2 when they use all DA absorption, and remained 2.4 after
they had removed metal lines and DLAs. They stated that this residual
scatter was probably cosmic variance.

We now measure the dispersion of DA values measured in segments of
spectra of length $\Delta z = 0.1$, and we estimate the individual
contributions of the low density IGM, the \lya\ lines of LLS and the
metal lines.

\subsection{DA2: The Mean DA in $\Delta z = 0.1$ Intervals}

We have already displayed the standard deviation of the pixel values
of DA, $\sigma (DA0)$, on various Figures, and we saw that the values
were large and similar in size to the mean DA values.  A significant
part of the $\sigma (DA0)$ values is from the photon noise.  However,
had we removed the photon noise, the $\sigma (DA0)$ values would be
smaller than the standard deviation of the true flux from the QSOs,
because the Kast spectra do not resolve spectra lines. The Kast
spectrograph makes absorption lines shallower.

We have examined the dispersion in the DA values using segments of
spectrum of length $\Delta z = 0.1$, which is 121.567~\AA\ in the
observed frame or 108 Kast pixels. The
velocity interval from one end of a segment to the other is
10335~\kms\ \citep[Eqn. 20]{sargent80}, and in a model with H$\rm
_o=71$~\kms~Mpc$^{-1}$, \ol $= 0.73$, and \om $= 0.27$, $\Delta z =
0.1$ centered at $z=1.9$ is 152.57 comoving Mpc.

We define a DA2 value as the mean DA in a segment of the spectrum of
one QSO that is 121.567~\AA\ long.

\begin{figure}
\epsscale{1.0}
\plotone{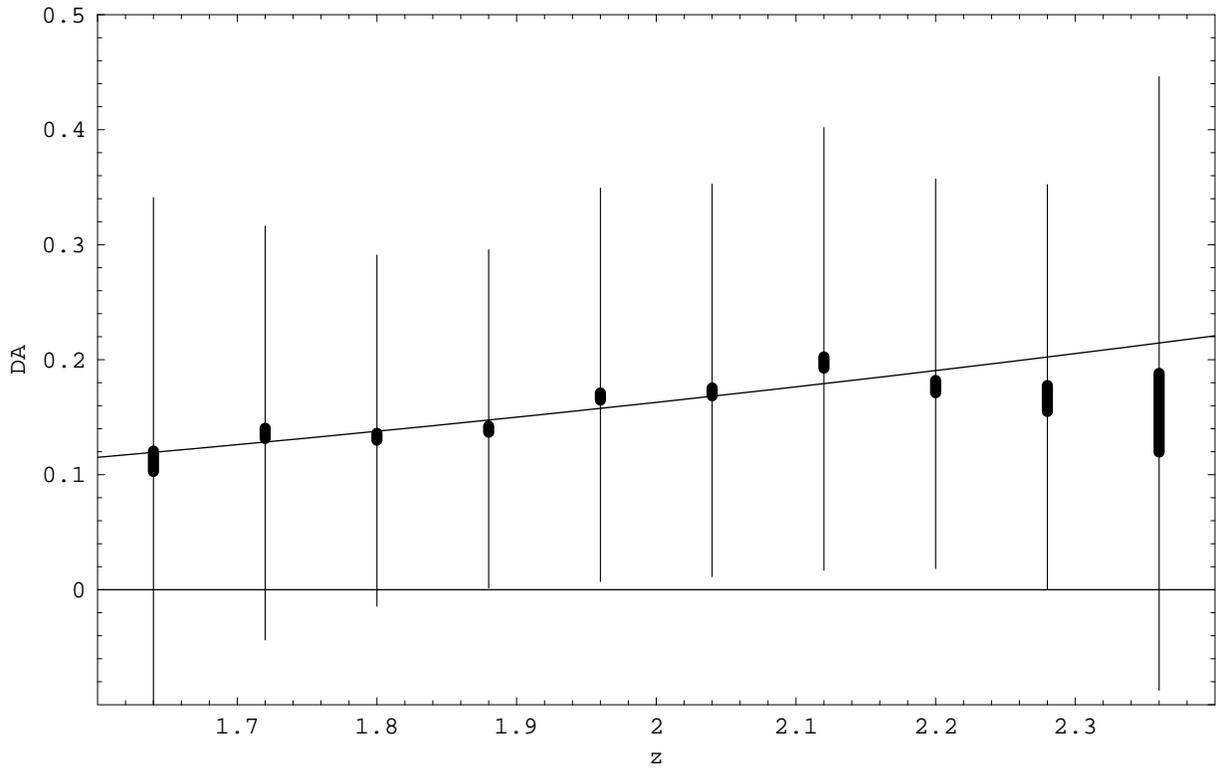}
\caption{\label{flx}
The total amount of absorption in the Kast spectra from rest
wavelengths 1070 -- 1170~\AA\ as a function of \lya\ redshift.  We
show the mean DA values for all pixels with \zabs\ in the indicated
ranges.  The measurements from each QSO contribute to 1 -- 3 bins. The
DA values include contributions from the \lya\ in the IGM, in LLS,
DLAs and metal lines. The errors are as in similar Figures.}
\end{figure}

\begin{figure} 
\epsscale{1.0} 
\plotone{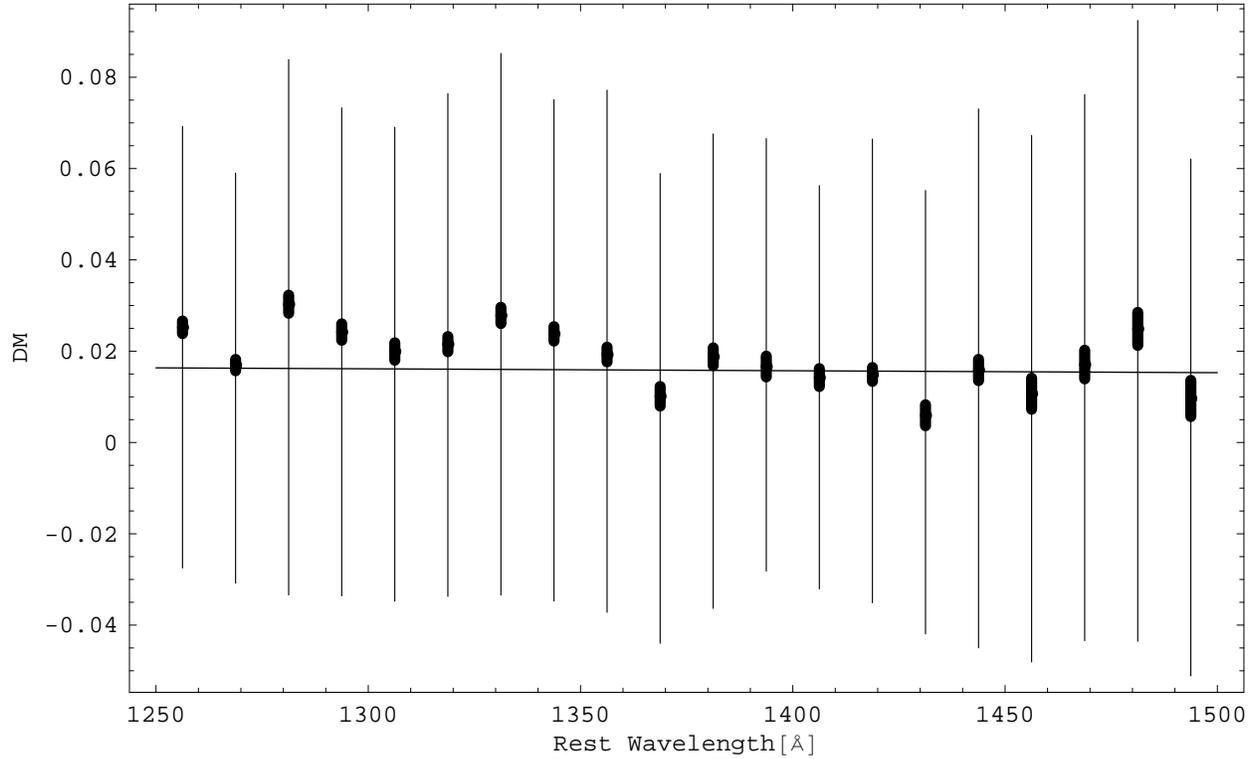} 
\caption{\label{fcx}
The amount of absorption in the Kast spectra from metal lines, DM, as
a function of rest wavelengths.  For our convenience, in this plot,
we use DM values that we calculated using the raw fitted continua,
without correction for the correlations with SDA or SNR2.  Because of
this, the points are too high by an average of 0.495\% and the mean
value on the plot is 2.36\%, instead of the correct value of
1.87\%. Values can be negative because of photon noise and continuum
fitting errors.  The line is the DM1 fit.}
\end{figure}

\begin{figure} 
\epsscale{1.0} 
\plotone{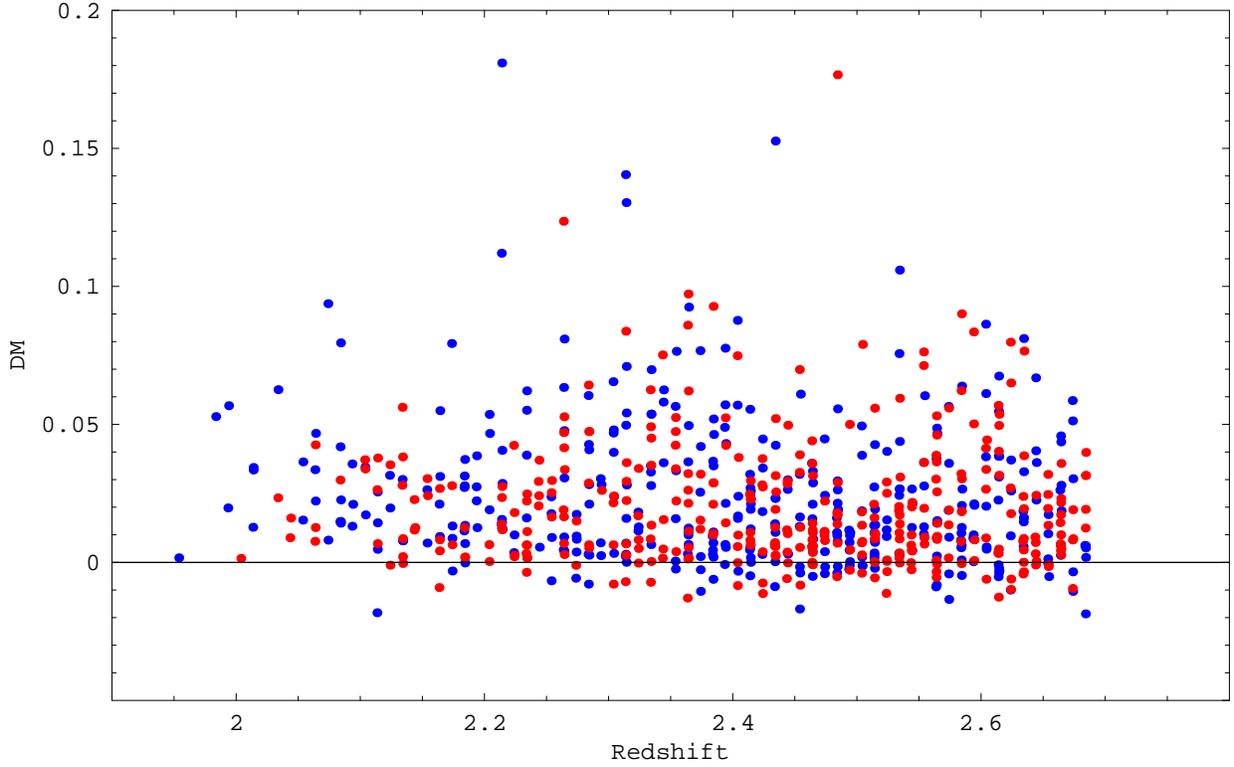} 
\caption{\label{dm2z}
The amount of absorption by metal lines in the Kast spectra, DM2
values, as a function of redshift $z$ for \lya .  Observed wavelength
is $1215.67 \times (1+z)$.  The $377 \times 2$ DM2 values are the mean
values, in adjacent segments of spectrum 121.567~\AA\ long.  The
lighter points (red) are from bins shifted from the darker (blue) ones
by $\Delta z = 0.05$, one half the bin size of both points, hence most
portions of a given spectrum contribute to two points.  For our
convenience, in this plot, we use DM2 values that we calculated using
the raw fitted continua, without correction for the correlations with
SDA or SNR2.  Because of this, the points are too high by an average
of 0.495\% and the mean value on the plot is 2.36\%, instead of the
correct value of 1.87\%. Values can be negative because of photon
noise and continuum fitting errors.  }
\end{figure}

\begin{figure}
\epsscale{1.0}
\plotone{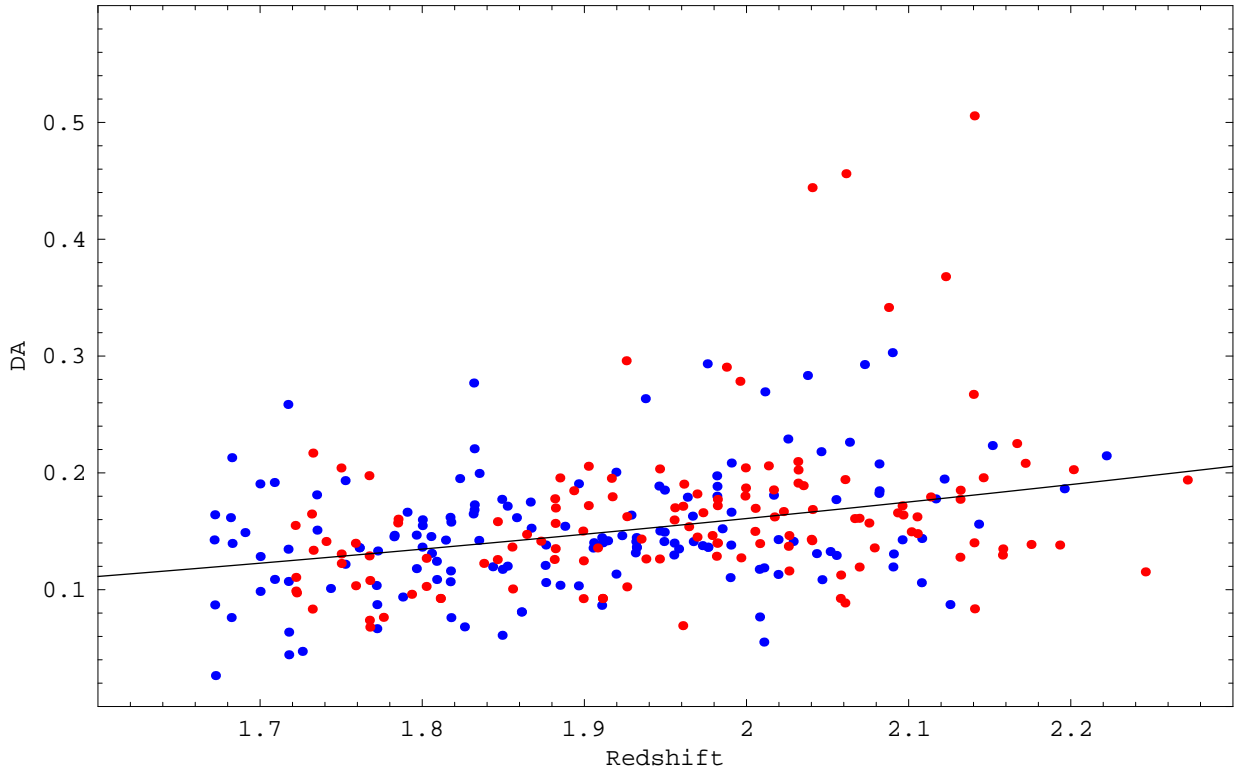}
\caption{\label{fc}
The total amount of absorption in the Kast spectra in the rest frame
interval 1070 -- 1170~\AA\ as a function of redshift of \lya .  The
absorption includes \lya\ from the IGM, LLS, DLAs and metal lines.
The points are DA2 values, the means in segments of length $\Delta z =
0.1$ or 121.567~\AA\ in the observed frame.  The solid line indicates
the fit to the redshift evolution of DA, given by Equation (4).  The
lighter points (red) are from bins shifted from the darker (blue) ones
by $\Delta z = 0.05$, one half the bin size of both points.  Most of
the largest values come from DLAs, some of which are split between two
adjacent bins.  }
\end{figure}

In Figure \ref{fc} we show the DA2 values from the Kast spectra.  We
started at the lowest $z$ covered by each spectrum, and made points
for each 0.1 segment, discarding the remainder of length $<0.1$ at the
high z end.  This explains why there are many dark (blue) points at
the lowest redshifts and few at the highest redshifts. We then added
light (red) points for bins starting at $z$ values larger by 0.05, to
better show the high redshifts and the DLAs. Many pixels then
contribute to two points on the plot.  We also show the DA(z) curve
from Figure \ref{flx}.

The contribution of photon noise to the $\sigma (DA2s)$ values is
small, except for a few segments.  The $\sigma $(DA2s) from photon
noise alone in a 108 pixel segment is 0.96\% for S/N $=10$ per pixel,
rising to 3.8\% by S/N $=5$.  The few segments with S/N $<5$ are all
in the lowest \zabs\ bin in Figure \ref{fc}, and they alone have
$\sigma$(DA2s) = 0.08, while the remaining segments have the
$\sigma(DA2s) = 0.06$. The same applies to the equivalent point on
Figure \ref{flx}.

Figure \ref{fc} shows that the distribution is asymmetric, with a small
fraction of segments having unusually large DA2 values. Due to the
accidents of the bin placement, 3 additional high points, all due to DLAs,
are seen in the shifted bins (red).  These DLAs are present in the
blue sample as well, but due to an accident of bin placement their DA
is split between two bins.  The dispersion of the (blue) points about
the mean DA($z$) (not the curve) is \datwosp \%, about half of the mean
DA due to \lya\ alone.

To remove the portion of the dispersion in DA that is due to the $z$
trend, we define DA2s, where the suffix ``s" refers to values
scaled to the values we expect at $z=1.900$ using the DA(z) trend from 
Equation \ref{daz}.

\subsection{Comparison of DA2s in Kast and Artificial Spectra}

\begin{figure}
\epsscale{1.0}
\plotone{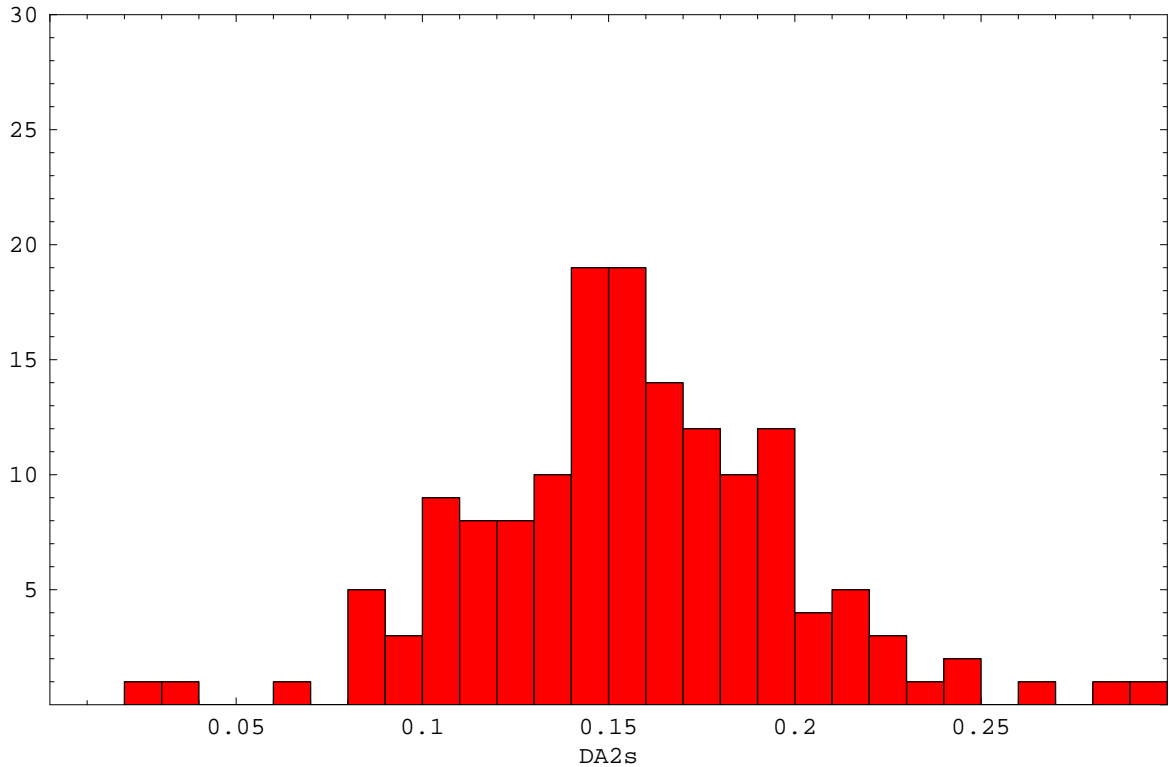}
\caption{\label{fmli}
The distribution of DA2s values from the Kast spectra.  These are the
mean DA values in bins of size $\Delta z = 0.1$. The values have been
scaled, pixel by pixel, to the DA expected at $z=1.9$ using the trend
of Equation (4).  These DA values include absorption by metal lines
and the \lya\ lines from the IGM, LLS and DLAs.  Each portion of a
spectrum contributes to a maximum of one value to this
histogram. Unlike Figure \ref{fc} we do not use bins shifted by 0.05 in
$z$.  One DA2s value is too large to appear on the plot.  }
\end{figure}

\begin{figure}
\epsscale{1.0}
\plotone{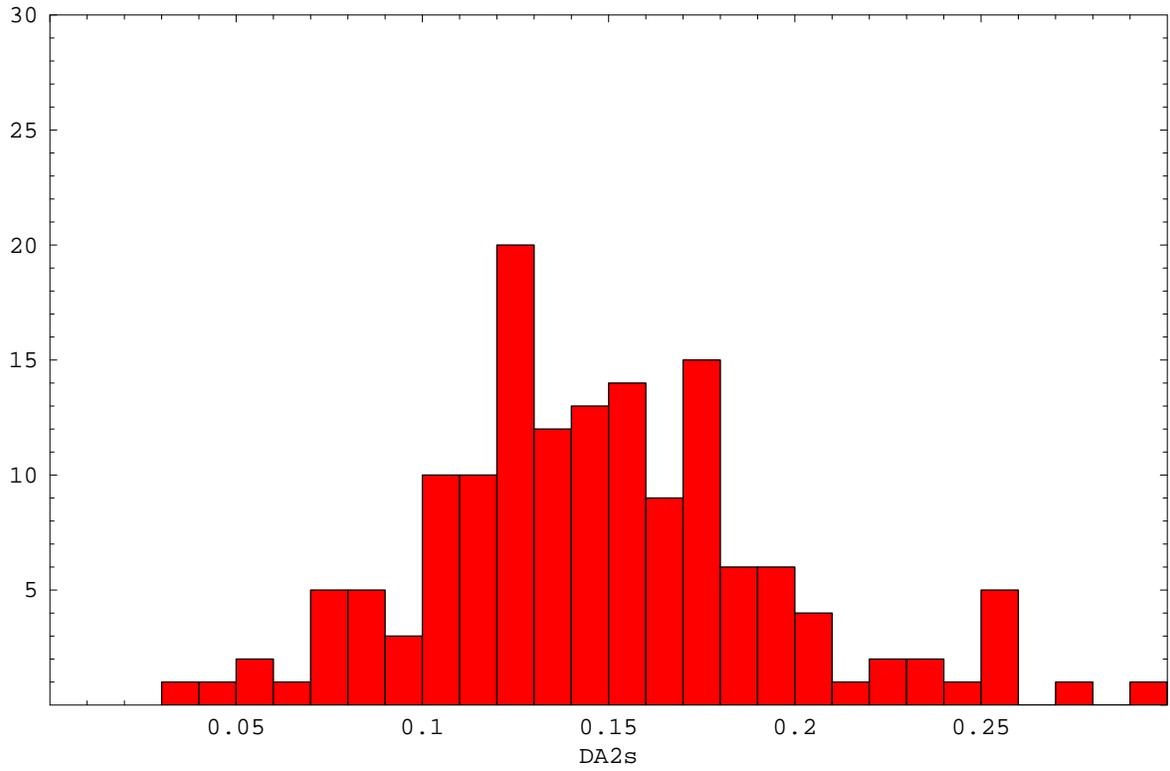}
\caption{\label{fmlii}
As Figure \ref{fmli}, but for the artificial spectra that we used to
determine the corrections we applied to our continuum levels.  All the
DA2s values are shown.  }
\end{figure}

In Figures \ref{fmli} and Figure \ref{fmlii} we show the distribution
of DA2s in the Kast and the artificial spectra. We list parameters of the
distributions in rows 1 and 6 of Table \ref{tabb}. 
We have not adjusted the artificial spectra
to exactly match the absorption in Kast spectra, especially at this
low $z$, and hence any agreement is accidental.  We see that the two
have similar mean DA2s, however the Kast have a larger dispersion, as we
might expect because they contain \lya\ lines from LLS and metal lines.
For example, the Kast spectra include three very high DA
values, all from DLAs, that do not show on Figure \ref{fmli}.  In row 2 of 
Table \ref{tabb} we also give statistics for just
the segments with DA2s $<0.35$.
Both the mean and the standard deviation are reduced as we would
expect, but the $\sigma $(DA2s) for the Kast segments remains larger
(about 3$\sigma$ level) than that for the artificial segments. 

The approximate similarity of the artificial and Kast spectra means
that the artificial spectra were well suited to help us find and
remove the biases in the continuum fits. The artificial spectra
accidentally contained about the correct total amount of absorption,
although they were designed to model just the low density IGM. If we were
attempting to make the artificial spectra very similar to the Kast
ones, they should have less absorption from the IGM, and they should 
include contributions from LLS and metals.

\subsection{Calculation of the dispersion in the DA from the IGM }

We can calculate the expected variation in the amount of absorption in
a spectrum. We restrict the calculation to the absorption from the
\lya\ lines in the lower density portions of the IGM. We ignore
absorption by the \lya\ lines of LLS and metal lines.  We shall
introduce the notation
\begin{equation}
\sigma (\Delta z ) = \rm \sigma (DA2s~low~density~IGM~only) 
\end{equation}
for the standard deviation of the mean DA in spectral segments of
length $\Delta z = 0.1$.  The \sigzo\ values are the 1D flux 
analogues of the 3D mass $\sigma
_8$ values, \citep[Eqn 9.18, Fig 9.2]{kolb90}.  The \sigzo\ value is
of great interest because it is a measure of the power in flux in the
\lyaf\ on scales of 153~Mpc, and it is related to the mass power
spectrum.

\citet[Fig. 10a]{mcdonald03} shows that the relationship between the
3D matter and flux power depend on scale.  On large scales the flux
power is proportional to the matter power, and hence to $\sigma_8^2$.
On small scales an increase in the matter power leads to larger
velocities that smooth the fluctuations in the flux, and decrease the
flux power.  At a scale of $k \simeq 0.013$ (s/km) (1.5~Mpc) the flux
power is insensitive to the mass power.

We can derive the variance of the DA2 values 
by integrating the power spectrum of the flux over a top hat window 
function of width $\Delta z =0.1$ .

First of all, we generate one dimensional flux power spectrum, 
$P_{F1D}(k)$, by following the \citet{mcdonald03} parameterization.
We introduce $\delta A_{1} = 0.218$ to adjust the cosmological
parameters and redshift, and $\delta \overline{F} = 0.0419$
to match the mean flux with our measured value
\citep[Table 1]{mcdonald03}.
Then we calculate,
\begin{equation}
\sigzo ^2 = \left< F\right>^2 \int P_{F1D}(k) W^2(k) dk
\end{equation}
where $W(kR)$ is the top hat window function.
We correct the result to our definitions of flux, since in
our units \citet{mcdonald03} uses $(F-<F>)/<F>$, a flux deviation
divided by the global mean flux at that $z$.  For one dimensional case, 
we have
\begin{equation}
W^2(kR) = { {2(1-cos(k R))} \over {(k R)^2} }
\end{equation}
where $R=$ 10335 \kms\ for $\Delta z=0.1$.

We obtain \sigzo $=0.057$ at $z=1.9$ with DA=0.118 and 
$\sigma _8 = 0.9$, for the low density IGM alone.

\subsection{Measurement of the dispersion in the DA from the IGM }

A measurement of \sigzo\ has several uses. It provides a check on the
linear model bias parameters $b_{\delta}^{'2}$ and $\beta $
\citep{mcdonald03}.  The scale of \sigzo\ is large enough that we are
well in the linear regime, and the astrophysical complications such as
fluctuations in the thermal history and temperature and UVB will be
less important than on small scales.
If we know the bias parameters from calculations, we can calculate a
normalization for the mass power spectrum that is a larger scale and 
earlier epoch analogue of of $\sigma _8$.

The \sigzo\ value is hard to measure because we must estimate and
subtract the portion of the dispersion coming from the \lya\ lines in
LLS, metal lines and continuum fit errors.

In Table \ref{tabb} we list the mean and $\sigma $ of the DA2 values for the
\lya\ lines of LLS (including DLAs).  We calculate that $\sigma $(DA2)
for the \lya\ of LLS is 0.035 by randomly sampling their $W_r$
values. We have checked this result analytically.  Since the expected
number of LLS per $\Delta z = 0.1$ is only 0.14, 0.869 of segments
have none, 0.122 have one, and 0.009 have two or more, if we ignore
clustering.  We then obtain the approximate $\sigma $ of the DA of the
\lya\ of the LLS from the sum of the squares of the $W_r$ values.  In
Table \ref{tabb} we also list the equivalent values for the metal lines alone,
mean DM2 and its $\sigma $ taken from the end of \S \ref{sfm}.

Both the \lya\ of LLS and the metal lines have a much larger effect on
the $\sigma (DA2)$ values than they do on the mean DA2 values.  The
standard deviations of the amount of absorption from the \lya\ of LLS,
$\sigma (DA2)$, is about 3.4 times the mean value.  Similarly, the
metal line $\sigma (DA2)$ is 1.3 times the mean.  Both distributions
have long tails with huge values.  The two distributions are more
asymmetric than either Poisson or exponential distributions, both of
which have $\sigma $ equal to their means.

We estimated the $\sigma $ of the DA2s of the \Lya\ in the IGM alone,
\sigzo\ by adding random contributions from the IGM, LLS and metals.
We assume that the DA2 values for the \lyaf\ alone are drawn from a
Normal distribution with mean 11.8\% and the one free parameter, the
unknown \sigzo .  We made mock DA2 values by adding together three
random numbers:
\begin{itemize}
\item DA2 from the \lyaf\ alone: A random deviate from the Normal distribution.
\item DA2 from the LLS: A randomly selected DA2 value drawn from the known 
distribution of $W_r$ values for LLS (including DLAs).
\item DM2 from the metals: A randomly selected DM2 value, scaled to 1120~\AA .
\end{itemize}
The mean DA2 for these mock spectra does not change as we vary the
\sigzo\ value. We found that the mean DA2 from the mock spectra was
0.151, about $1\sigma $ lower than the mean DA2s that we give in Table
\ref{tabb}, and identical within the errors to the value in Table
\ref{taba}, as expected, since we are using the same data.  We
estimate a 68\% confidence interval of 
$\rm 0.032 < $ \sigzo\ $ < 0.044$, 
with a best value of 0.039. For these values, the mock
segments matched the observed $\sigma (DA2s) = 0.0612 \pm 0.0035$.

The values we give for \sigzo\ may have a larger error than we have
estimated, because the many factors that contribute to the measurement
are not well known. The dispersions from the metals involves an
extrapolation, and we have assumed that the contribution from the
continuum error is insignificant.

The \sigzo\ value includes the dispersion from both the \lya\ in the
low density IGM and the continuum errors, because we did not include
the latter in three part model for the mock segments.  
The dispersion of the DA in
the artificial spectra, without noise or continua fits, was 0.0441.
When we re-measured this dispersion after adding photon noise and
fitting the continua we found 0.0422, from Table \ref{tabb}, which shows no
significant increase from the fits.  This implies that the continuum
fits have an insignificant error.  Below we will see that the
dispersion from a hydrodynamic simulation suggests that our \sigzo\
value is approximately correct, and hence that the continuum error is
not a major part of the value.  However, if we treat the continuum
fits as absorbed spectra, the dispersion of the equivalent DA is
0.0325, which suggests that the continuum error is a significant part
of the \sigzo\ value.  We intend to investigate this issue in future
work starting with a fully automated continuum fitting algorithm that
should be more stable.

The \lya\ absorption from the low density IGM, the \lya\ from LLS and
the metal lines all appear to contribute similar amounts to the
dispersion in the DA2s values in $\Delta z =0.1$ bins at $z=1.9$.
\citet[Fig. 3]{kim04} showed that metal lines comprise a large part of
the power on small scales ($k > 0.05$ s/km), but only 0.07 of the
total power on scales $\simeq 100$~\kms . Since integrated power is
proportional to variance, this implies that the metals will contribute
about 7\% of the variance on scales of 100~\kms , which is only 1\% of
the 10,335~\kms\ span of $\Delta z =0.1 $.  We find that the metals
comprise $(0.031/0.0612)^2 = 0.26$ of the total variance sampled in
bins with $\Delta z = 0.1$.  On these large scales the dispersion due
to the metals becomes a larger part of the total because the
absorption by metal ions is strongly correlated across these large
scales.  A single absorption system can put strong metal lines all
over a spectrum. The correlation arises from the multiplicity of
strong spectral lines, and not from the distribution of matter on
153~Mpc scales.

Our \sigzo\ value may be too large because we ignored the correlation
of LLS and metal lines.  The absorption systems that show many strong
metal lines are likely to be LLS. The DLAs in particular also have a
lot of metal lines, especially when the gas covers a wide range of
velocities, and the metal abundances are high.  Hence the mock spectra
that we made to calculate \sigzo\ would be more realistic if the we
had chosen a large DM2 value, rather than a random one, whenever a segment
had a \lya\ with a large $W_r$ value. This correction would make more
segments with very large DA2 values, which would increase the variance
from the metals and LLS, and hence decrease our estimate of the \sigzo
.  We have not modeled this complication, because the metal lines lie
at specific velocities relative to the \lya\ line and we should make
complete mock spectra, with realistic mock absorption systems.
Alternatively, and preferably, we could use spectra of high spectral
resolution and remove the metal lines and the \lya\ of the LLS.

We have no good explanation of why the \sigzo\ value that we measured
, $0.039^{+0.005}_{-0.007}$ is $3.6 \sigma $ less that the value we
calculated. Measurement error is a clear option, especially 
since this is the first ever such measurement.
However,systematic effects from the photon noise, continuum fit and
LLS -- metal correlations should all have lead our measured value too
be large, not too small, and to obtain a \sigzo\ = 0.057, we would have 
to have greatly overestimated the $\sigma $ from both the metals and the LLS.
The calculation is also uncertain and untested and involves various uncertain
scalings.

\section{DA3: The DA in Quarters of the Sample}

To obtain an accurate mean DA value we need to average over a
substantial path length, from many lines of sight, because the
distribution of DA is not normal.  If the DA0 values were normal
deviates we would expect that the standard deviations of the DA2s
points would be about 10 times smaller than that of the DA per pixel,
DA0, because there are about 108 pixels per DA2 point.  However
$\sigma $(DA2s)$=$\dastwosp \%, only 3 times less that $\sigma $(DA0s)
= \dasZerosp \%.  When we average from 1 to 100 pixels the DA0
distribution is asymmetric because the \lyaf\ at $z=1.9$ is largely
absorption free, with well separated lines making most of the
absorption.  \citet[Appendix C]{bernardi03} discuss another
consequence of this asymmetry.

We have explored the dispersion of the mean DA in much larger bins.  We
divided the spectra, both artificial and Kast, into four groups, or
quarters, giving the DA3s values that we list in Table \ref{taba}.  We
used Equation (4) to scale these values to the expected value at
$z=1.9$, because the fourth quarter happened to have a significantly
lower mean $z$.  We assigned the QSOs to quarters in RA order, to
emphasize a possible source of bias. QSOs with similar RA values are
more likely to be observed at the same time, through similar
conditions, giving more similar S/N, and to have their spectra reduced
in the same way. Our initial continuum fits were also done in RA
order, although later adjustments were done in order of emission line
strength, and S/N.  The RA order of the artificial spectra is that of
their partner QSOs.

The first column is the DA3s in the input artificial spectra, with
flux range 0 -- 1, plus photon noise.  If the simulations were a
faithful representation of the IGM absorption, the error on the mean
of these values, $\sigma (\mu ) = \sigma /2 = 0.48$\% would be an
estimate of the portion of the error in the DA from the 77 Kast
spectra due to the sample size.  From an examination of a much larger
sample of similar artificial spectra, we calculate an expected $\sigma
(\mu ) = 0.4 - 0.5$\% for a sample at $z=1.9$ with a path length of
19.75.

This is the error we would have obtained had our spectra been perfect,
and the continuum fits perfect. However, the artificial spectra have
more absorption than the low density IGM alone, and they lack high
\nhi\ lines and metal lines. Hence the error for the Kast spectra will
be somewhat larger.

The second column shows our measurements of the DA3s in the artificial
spectra with emission lines and S/N like our Kast spectra.  The
dispersion in this column will include many effects, especially the
continuum level errors and our corrections for the biases that we
measured as a function of SNR2 and SDA.  The mean of this column is
nearly identical to the equivalent values in the first column because
the corrections that we made to the continua forced agreement.  The
$\sigma $ values are similar because the continuum fits did not
increase the dispersion.

The third column shows the row by row difference between the DA3s
values in the first two columns: the four true DA3s values, minus our
measurement of each.  The dispersion in this column is the portion of
the measurement error coming from the S/N, and especially our
continuum fitting and continuum level corrections.  The error on the
mean of $\sigma (\mu ) = 0.32$\% is again less than the error in the
first column, $0.48$\%.  This implies that our spectra, their S/N, our
continuum fits, and our corrections to the continuum fits are adequate
for a sample of 77 QSOs at $z=2$.

The fourth column shows the DA3s values for the quarters of the Kast
QSO spectra. The error on the mean is the largest of the values for
the different columns, as expected, because we are now sensitive to
the large \nhi\ lines and metal lines. We already saw working with the
DA2s $\Delta z =0.1$ values that the metal lines and the \lya\ of LLS
have a large effect.

\section{Error on Our DA Measurement}
\label{error}

The $1\sigma $ error terms associated with our DA4 estimate are:
\begin{itemize}
\item 0.64\% estimated from the dispersion in quarters of our sample
(Table \ref{taba}, column 5).  This includes sample variance and that
part of the calibration error that varies between the quarters.
\item 0.32\% calibration error, from Table \ref{taba} column
  4. Estimate of the error in the corrections that we applied to the
  continuum levels that correlated with SNR2 and SDA.
\item 0.48\% sample variance, from the size of our sample, (sometimes
  called cosmic variance).
\item 0.6\% uncertainty in the amount of metal absorption 
\item 0.4\% uncertainty in the amount of absorption from \lya\ lines
  with \lnhi $> 17.2$~\cmm . 
\end{itemize}

To obtain DA4s, we first scaled the DA0 values for each pixel to the
expected value at $z=1.9$, using Equation (4), before taking the mean.
The result is slightly different from a scaling of the mean DA4, using
the single mean \zabs\ value.  As our estimate of the error on our
DA4s measurement we use the quadratic sum of the errors on the first
two terms above, giving
\begin{equation}
DA4s(z=1.9) = 15.1 \pm 0.7\%,
\end{equation}
where the suffix ``s" means that the value applies to $z=1.900$.

For comparison with measurements or simulations that do not include
metal lines, we can subtract $DMs = 2.3 \pm 0.5$\% for the metal
lines, giving DA7s = DA4s - DMs, or
\begin{equation}
DA7s(z=1.900) = 12.8 \pm 0.9\%,
\end{equation}
and for comparison with data or simulations that include neither metal lines 
nor
\lya\ lines from systems with \lnhi $> 17.2$~\cmm\ (all LLS and DLAs) the
appropriate DA is DA8s = DA4s - DA6s - DMs, or 
\begin{equation}
DA8s(z=1.900) = 11.8 \pm 1.0\%.
\end{equation}

\section{Comparison with Hydrodynamic Simulations}
In this section we will compare the DA from the Kast spectra to values
from full hydrodynamic numerical simulations.
We will give a more thorough description in Jena \etal\ 2004.

We used the cosmological simulation code, ENZO \citep{norman99},
which follows both dark matter dynamics and hydrodynamics consistently. The
collisionless dark matter particles are evolved using a Lagrangian
particle-mesh method whereas the equations of gas dynamics are solved
using a piecewise parabolic (PPM) method.

The simulations assume an evolving UV background radiation field due
to both quasar and stellar sources as described by \citet{madau99}.
From this spectrum we calculate photo-ionization and photo-heating
rates for H I, He I, and He II assuming the gas is optically thin.
Although we believe that we have the appropriate level of ionization
in the low density IGM, the gas remains optically thin in the
simulation even in high density regions, and hence we over ionized the
densest regions and the spectra do not include any LLS or DLAs.  The
highest column density in any of the artificial spectra that we made
was \lnhi $<17$~\cmm .  To account for the extra heating from late
He~II reionization due to opacity effects \citep{abel99}, we multiply
the He~II photo-heating rate by $\gamma _{228} = 1.8$.  \citet{bryan00}
have shown that this level of heating is needed to match the
$b-$parameter distribution of the \lyaf . This is equivalent to $\gamma
_{228} \simeq 2.4$ in simulations with higher resolution.

We used a spatially flat $\Lambda$CDM universe with $\Omega_b =
0.044$, $\Omega_m = 0.27$ and ${\Omega_{\Lambda} = 0.73}$.  We used
H$\rm _o=71$~\kms~Mpc$^{-1}$ an initial power spectrum index $n= 1.00$
and we normalized the power spectrum to ${\sigma_8 = 0.9}$. These
values are similar to the best fit values using the WMAP first year
data \citep{spergel03}, but we chose a slightly larger $\sigma_8 $
consistent with other measurements.

We ran the simulation with a box size of $54.528 h^{-1}$~Mpc, or
75.7 comoving Mpc with a grid size of ${1024^3}$.  This provides us
with enough resolution and included the effects of some of the larger
scale power missing from smaller boxes. The initial perturbations were
assumed to arise from a Harrison-Zel`dovich power spectrum with a CDM
transfer function. These density perturbations were then converted to
initial velocities using the Zel`dovich approximation. The simulation
was run on the Blue Horizon computer at the San Diego Supercomputing
Center.

We make several points about the comparison of the hydrodynamic
simulation and the Kast spectra.

First, we should compare the simulated spectra to the DA values for
the IGM only, without metal line or the \lya\ of LLS.

Second, although the DA in a single DA2s segment will differ from the
mean in the universe, by design the mean DA in the simulation box
should be the mean in the universe for the physics and parameters that
were simulated.

Third, the 150 simulated spectrum segments were all made by passing
lines of sight through the same 76~Mpc volume. To obtain one segment,
the line of sight traveled about two times through the box.  Their
dispersion is reduced because the same modes, both short and
especially long ones, are sampled many times.

Lastly, the simulation was made in a 76~Mpc box that contained no
power on scales $\geq 38$~Mpc, and reduced power on somewhat smaller
scales because the boundary conditions were periodic.

\begin{figure}
\epsscale{1.0}
\plotone{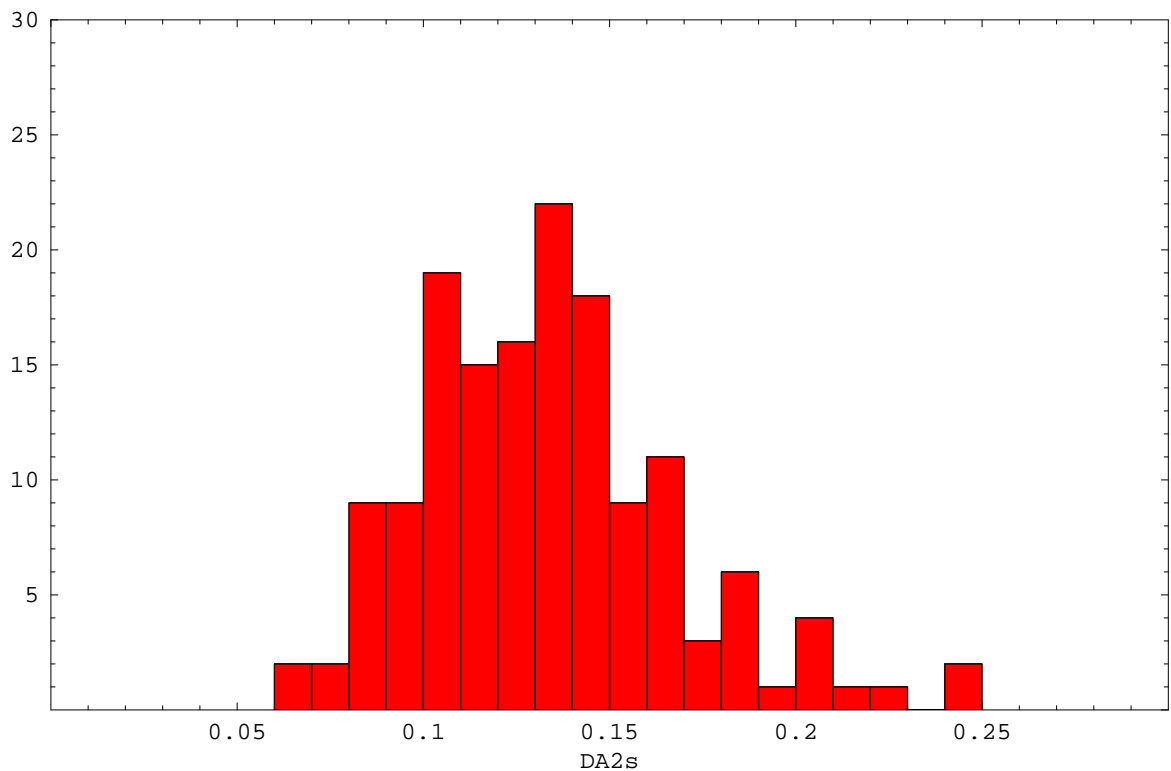}
\caption{\label{fmlv}
As Figure \ref{fmli}, but showing DA2 values for spectra from a full
hydrodynamic simulation of the IGM, in a 75.7~Mpc with a grid size of
${1024^3}$. The simulation does not include \lya\ from LLS and DLAs,
and it does not include metal absorption.  The bins were taken between
$1.9 < z < 2.0$, and the simulation evolved slightly in this
interval. We did not scale these DA2 values to $z=1.9$.  }
\end{figure}

We list the results from the spectral segments of length $\Delta z =
0.1$ from the hydrodynamic simulation in row 7 of 
Table \ref{tabb}, along with the
other comparable measures of the DA2s.  
In Figure \ref{fmlv} we show the distribution of DA2s from the
hydrodynamic simulation. The mean $DA2s = 12.87 \pm
0.27$\% is $1.08 \pm 0.09$ times that for the Kast IGM only from Table
\ref{tabb}. The simulation had slightly too much absorption.  The $\sigma
$(DA2s) of the simulated spectra was $0.85 \pm 0.14$ times the \sigzo\
from the Kast IGM only, slightly too small.  We expect the simulations
to have the same mean and a smaller $\sigma $ than the Kast IGM only
values.

There are three obvious ways to adjust the parameters of the
simulation to better match the DA of the Kast spectra.  We could
increase the UVB intensity, which reduces the amount of H~I. We
could reduce the \obh\ which has the same effect, or we could increase
the $\sigma _8$ which removes baryons from the low density IGM where
they absorb most.

\begin{figure}
\epsscale{0.7}
\plotone{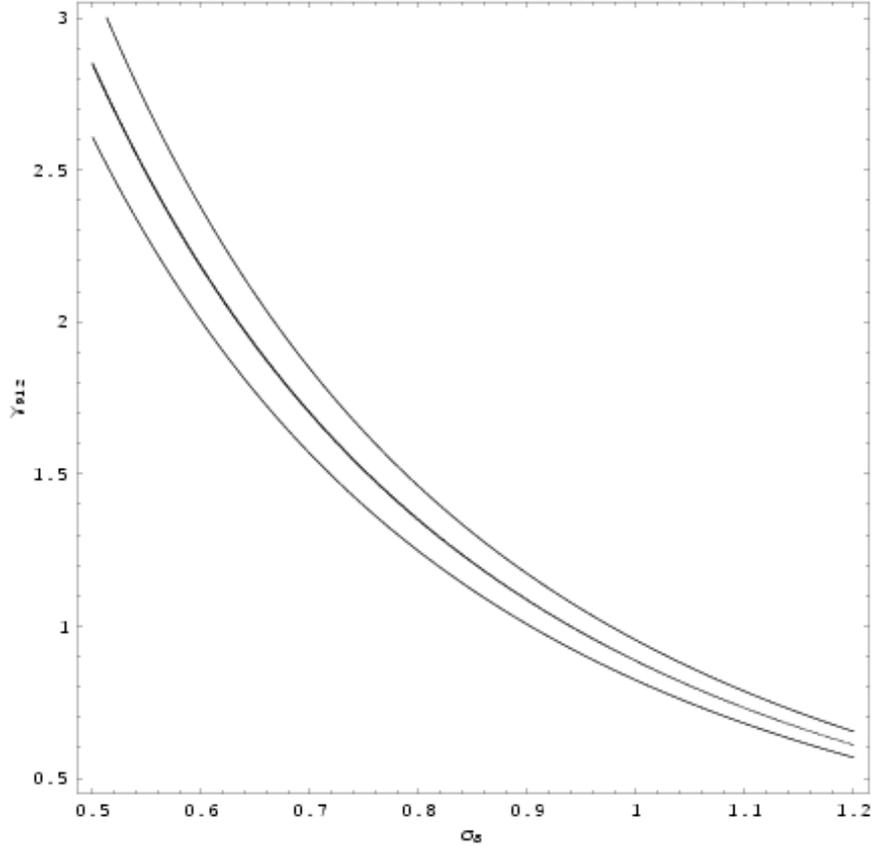}
\caption{\label{gammasig}
The ionization rate per H~I atom in the IGM as a function of the
amplitude of the matter power spectrum $\sigma _8$.  The vertical axis
is the ionization rate per H~I atom, $\Gamma $, in units of that given
by \citet{madau99}. When we adopt their spectrum $\Gamma $ is
proportional to the intensity of the radiation intensity $J_{912}$.
The central curve shows the approximate $\Gamma $ and $\sigma _8$
values that give the $ DA8s =0.118 \pm 0.010$ that we measured in the
Kast spectra for the low density IGM only, at $z=1.9$.  The outer two
curves give the $\pm 1 \sigma $ range for the DA8s.  These DA values
exclude metal lines and the \lya\ lines of LLS.  Models with larger
$\Gamma $, and larger $\sigma _8$ than the curves, in the upper right,
have too little absorption, while those below have too much.  Larger
$\Gamma $ values leave less H~I and less absorption than we
see. Larger $\sigma _8$ leaves fewer baryons in the IGM where we get
the most absorption per baryon.  }
\end{figure}

In Figure \ref{gammasig} we show the approximate intensity of the UVB required
to explain the mean DA at $z=1.9$.  The vertical axis $\Gamma $ is the
photo-ionization rate per H~I atom in the low density optically thin
IGM. We use the units of the predicted rate at $z=1.9$ from
\citep{madau99}:
\begin{equation}
\Gamma = 1.329 \times 10^{-12}\gamma _{912} \rm~s^{-1},
\end{equation}
where $\gamma _{912} $ is a dimensionless number.  \citet{madau99}
predicted $\gamma _{912}=1$, and when we adopt their spectrum shape,
the ionization rate is proportional to the intensity of the UVB, $J$.
We show how the UVB intensity required to give the $DA8s = 0.118 \pm
0.010$ for the Kast IGM only as a function of $\sigma _8$. We found
these curves using scaling relationships that we derived from many
simulations that we ran with a variety of parameters in boxes four
times smaller.  A smaller intensity is required when we have a larger
$\sigma _8$, since there are then fewer baryons in the IGM that we
need to ionize.

The changes required in any one of the three parameters to match the
observed DA are modest.  In Table \ref{tabe} we list these changes. In
the first row we give published estimates for \ob\ and $\sigma _8$
from \citet{spergel03}, $\gamma _{912}$ from \citet{madau99}, and DA8s
from this paper. The simulations shows that these values are not quite
concordant.  We will discuss $\gamma _{228}$ elsewhere, since high
resolution spectra are needed to assess the changes it makes to the
widths of \lyaf\ lines.  In the second row we give the DA from the
simulation, which differs from the Kast value. The third and fourth
rows show the $\sigma _8$ and $\gamma _{912}$ values from 
Figure \ref{gammasig}.  We obtain the \ob\ estimate in the fifth row using
\begin{equation}
\Omega _b = 0.044 (\tau_{eff} / 0.1378)^{1/\alpha}
\end{equation}
from Equation 1, where $\tau_{eff} = 0.1378$ corresponds to the DA(z=1.9)
$=12.87$\% from the simulations with \ob = 0.044, and we use $\alpha = 1.7$.
The values in rows 2 -- 5 give alternative concordant models for the IGM.

The errors that we quote in rows 3 -- 5 of Table \ref{tabe}, for $\gamma
_{912}$, $\sigma _8$ and \ob\ are from the DA8s error alone, assuming
that all other parameters are known without error, and that the
simulation is an accurate representation of the IGM.  These errors are
as small or smaller than those usually quoted for these parameters,
which shows that the accuracy that we have obtained for DA is
sufficient to give new cosmological information.  For example, if the
$\sigma _8$ and $\gamma _{912}$ values had insignificant errors, the
\ob $= 0.0417 \pm 0.0022$ from our DA measurement would be more
accurate than that using our measurements of D/H: \ob $=0.042 \pm
0.004$ \citep{kirkman03}, and comparable in accuracy to that from the
first year WMAP data, \ob $= 0.0444 \pm 0.0018$ \citep{spergel03}.

Our DA value gives $\gamma _{912}$ to higher accuracy than has been
possible before. In the last row of Table \ref{tabe} we give $\gamma _{912} =
1.08 \pm 0.27$ or $\Gamma = (1.44 \pm 0.36) \times 10^{-12}$~s$^{-1}$,
where the error now includes the contributions from the errors on the
other three parameters that we list. The main contribution to this
error is from the uncertainty in $\sigma _8$. We illustrated this in
Figure \ref{gammasig}.

The DA value alone leaves a strong degeneracy between the best fit
UVB intensity and $\sigma _8$ values. 
We can break this degeneracy using the variation in the DA,
from the power spectrum of the flux, or from \sigzo\ if we know the 
linear bias for the appropriate model.

If we change the parameters to reduce mean DA, the $\sigma $(DA2s)
will also decrease, and it is already smaller than the Kast IGM
value. However, we expect the simulation $\sigma (DA2s)$ to be smaller
than the Kast, because of the finite size of the simulation box, and
it is beyond the scope of this paper to determine whether the reduced
value would be compatible with the spectra.  However, we did check
that the power of the flux in these lines of sight does approximately
match that in HIRES spectra on much smaller scales, $0.008 < \log{k} <
0.08$ (s/km) at these redshifts.

\section{Sample Size, Spectral Resolution, LLS and Metals}

The asymmetric DA distribution extends to scales $> 153$~Mpc.  There
remain correlations in the \lya\ absorption from the IGM, from the matter
power spectrum on large scales. The metal
lines are correlated because one system can create lines all over a
spectrum. The DA distribution is also asymmetric because the \lya\
lines of LLS and DLAs are rare events that produce huge DA values.
When we take the mean of 152 DA2s values from Table \ref{tabb}, we
would expect the $\sigma = 0.0612/\sqrt {152} = 0.50$\% if the DA2s
were uncorrelated and normally distributed.  Instead we find $\sigma
(DA3s) = $0.64\%, from Table \ref{taba}.  The $\sigma $ decrease more
slowly than $(\Delta z )^{-0.5}$.

To help us compare measurements from different samples, we can use a
first order estimate
\begin{equation}
\sigma (DAs, z=1.9) = 0.024(\Delta z)^{-0.4},
\end{equation}
where $\Delta z$ is the path length in the sample at $z = 1.9$, and
the coefficient and power are from a straight line fit to the
$\sigma(DA3s)$ from the quarters of the Kast the sample (Table \ref{taba}) and
$\sigma (DA2s)$ from the $\Delta z = 0.1$ (Table \ref{tabb}).  This $\sigma $
is for DA values that include metal lines and the \lya\ of LLS.  Each
QSO contributes a maximum of $\Delta z = 0.156(1+z_{em})$ when we use
all wavelengths from \lyb\ to \lya , and, for our sample, typically
around 0.6 of this or $\Delta z = 0.3$ at $z=1.9$.

We expect to obtain approximately the same $\sigma (DA)$ with four
times smaller $\Delta z$ when we use spectra that are free of both the
metal lines and the \lya\ from LLS. Our estimate that the $\sigma
(DA2s) = 0.039 \pm 0.006$ implies that we might achieve
\begin{equation}
\label{siglyaonly}
\sigma (DA) = 0.012(\Delta z )^{-0.5},
\end{equation}
with such spectra.  This suggests that we might obtain an error on
DA8s (IGM only) of 1\% with only 5 high resolution spectra, comparable
to the error we obtained with the 77 Kast spectra, after we removed
the mean absorption by LLS and metal lines.  However, we would need
improvements in the flux calibration and continuum fitting to echelle
spectra to obtain errors in DA of 1\%. \citet{suzuki03b} found it very
difficult to obtain such small errors, using purpose built software,
and with ample calibration spectra.

\section{Comparison with Prior Measurements}
The few prior estimates of DA at $z=1.9$ all involve small samples and
they are all compatible with our new measurements.

\citet[Figure 15]{kim01} show measurement from about 11 QSOs near that
$z$, with a mean DA of about 10\% and a range of 6\% -- 15\% that
includes the values from the Kast spectra. They fit a power law to
\taueff\ values from UVES and HIRES spectra, some of which include
systems with high \nhi\ lines, but no metals, giving DA $=10.9$ at
$z=1.9$. Our equivalent value, between the values for DA7s and
DA8s, is approximately 12\%.

\citet{rauch97} measured DA in 7 HIRES spectra.  They identified and
rejected metals and all lines with $b<10$ \kms .  At $z=2.0$ they
found DA $14.8$\%, with no error offered.  Since only two QSOs
contributed data at $z < 2.3$, we estimate their error is $>1.6$\%
(Equation \ref{siglyaonly}).  Our equivalent value, DA7s $=12.8 \pm
0.9$\%, is smaller.

\citet{schaye03a} present a measurements of DA from high resolution
UVES and HIRES spectra of 19 QSOs, with 6.6~\kms\ resolution, 8 of
which contribute at $z=1.9$. Their best fit (Fig. 1) give DA(1.905)
$=12.6$\%, or 10.9 after they remove metal lines and \lya\ lines from
systems with \lnhi $> 19$~\cmm . Our equivalent value is approximately
$12 \pm 1$\%, and it is larger, but not significantly.

\citet{meiksin03} looked the apparent discrepancies in the estimates
of the average Ly$\alpha$ absorbed flux for $z>2.4$, and found that
when the results were interpreted on a consistent statistical basis,
all the estimates roughly agreed.  They find DA=$0.18 \pm 0.002$ at
$z=2.41$, which is basically consistent with the value that we would
infer at the same redshift (Figure 12).

\section{Discussion and Summary}
\label{si}

We have measured the amount of absorption in the \lyaf\ in spectra of
77 QSOs from the Kast spectrograph on the Lick 3m telescope.  We
measured the mean amount of absorption and the contributions from the
\lya\ lines of LLS and metal lines. We also measured the variance in
the amount of absorption from the metals, the LLS and the \lya\ in the
lower density IGM.  The amount of absorption that we find is
consistent with that in a large hydrodynamic simulation that uses
popular values for the cosmological and astrophysical parameters. We
summarize this work under these three topics, and the opportunities
for improvements.

\subsection{Mean DA}

We fit continua to the Kast spectra, and to artificial spectra that we
made to mimic them. The relative error in our continuum fits to the
artificial spectra is 3.5\% on average. The mean error for all 77
spectra is within 1--2\% of the correct value, except when the S/N per
1.13~\AA\ pixel is $< 6$ where we systematically placed the continuum
too high. We corrected this systematic bias and also our tendency to
place the continuum 0.5\% too high where there is a lot of absorption.

We find that the total absorption in the \lyaf\ between 1070 and 1170
\AA\ at $z=1.9$ is DA $=15.1 \pm 0.7$\%, including absorption by metal
lines and the \lya\ lines of LLS (defined to include all DLAs).  This
is the first measurement of DA to be made at any $z$ using a
calibrated continuum fitting procedure, and the first of any sort
using a large sample at $z \simeq 2$.

We measured the mean absorption due to metals at 1225 -- 1500~\AA\ in
both our 77 Kast spectra and from the lists of absorption lines in 26
spectra in \citet{sargent88a}.  The results agree. Near \lamr $=
1360$~\AA\ DM3 $= 1.87 \pm 0.13$~\% from our Kast spectra and DM4 $=
1.67 \pm 0.22$~\% from \citet{sargent88a}.

We must extrapolate the DM to obtain the metal absorption in the
\lyaf\ and this increases the uncertainty.  The extrapolation gives
$DM1-6(1120) = 1.92 \pm 0.42$\%, and we increase this by a factor of
1.2 to account for extra metal lines in the \lyaf , giving $DMs = 2.3
\pm 0.5$\%.

The total absorption in rest wavelengths 1070 -- 1170~\AA\ comprises:
$DA6s = 1.0 \pm 0.4$\% from LLS, $DMs = 2.3 \pm 0.5$\% from metals and
$DMA8s = 11.8 \pm 1.0$\% from the \lya\ in the low density IGM that
excludes \lya\ lines with \nhi $>17.2$~\cmm .

The absorption from metals is important.  At $z=1.9$ the metals are
$15 \pm 4$\% of the total absorption, and $19 \pm 5$\% of the
absorption by just the \lya\ with \lnhi $< 17.2$~\cmm .  We have
calculated the amount of absorption due to C~IV alone, which increases
slowly as \zabs\ drops, and hence \lamr\ for a given QSO sample.  By
\lamr $= 1120$~\AA\ at $z=1.9$ C~IV gives DM7 $= 0.80$\% that is 35\%
of the metal line absorption.

We calculated the absorption by the \lya\ lines of LLS, using a list
of rest equivalent widths from other Kast spectra, and normalizing to
the LLS density seen in HST spectra. We find DA6s $=1.0 \pm 0.4$\%,
where the error is nearly all from the uncertain density of LLS at
these low redshifts and the suffix ``s" refers to a value for $z=1.9$.
We calculated that the DLAs alone have DA $= 0.85 \pm 0.17$\%, which
is a larger proportion of the DA6s for all LLS than we
expected. Perhaps the DA6s value is too small.

\subsection{Dispersion of DA values}

We have measured the dispersion in the mean DA on large scales.  We
defined DA2s to be the mean DM in segments of a spectrum of length
$\Delta z = 0.1$, or 121.567~\AA\ in the observed frame.  This is
153 comoving Mpc at z=1.9. We scale the measurements to the amount
of absorption expected at $z=1.9$ since the evolution is significant.
We find $\sigma (DA2s) = 6.12 \pm 0.35 $\%.  This is a measure of the
amount of power in the flux distribution on scales similar to 153~Mpc,
however it includes the power due to the \lya\ of LLS and the metal
lines, and the power in the error in the continuum fits.

We have estimated the dispersion in the DA2s from the \lya\ lines of
LLS by making mock spectral segments of length $\Delta z =0.1$. We
added random samples of measured $W_r$ values to the segments. We
find $\sigma (DA2s, LLS) = 3.5 \pm 0.5$\%, much larger than the mean
value of $1.0 \pm 0.4$\%.  We also derived this $\sigma $
analytically.

We have measured the dispersion in the amount of metal absorption in
the Kast spectra: $\sigma (DM3) = 2.5$\%.  We found a similar value
from the \citet{sargent88a} spectra: $\sigma (DM4) = 2.7$\%.  These
are the measured standard deviations of the DM values in spectral
segments of $\Delta z = 0.1$, at $z=1.9$ and \lamr = 1360~\AA , which
is near \lamo = 4130~\AA .

We have estimated the dispersion that we expect from the metal lines
in the \lyaf . We scale from $\sigma (DM3, \lambda _r =1360) = 2.5$\%,
to obtain $\sigma (DMs)=3.1$\%. Since this value is from an
extrapolation, its error is large and not well known. It appears to be
larger than the mean DMs, $2.3 \pm 0.6$\%.

We calculated the dispersion of mean DA in segments of spectra
121.567~\AA\ long, from the absorption in the low density IGM alone,
is \sigzo = $3.9 ^{+0.5}_{-0.7}$\%. This value includes the dispersion
from the error in continuum fits which is probably small.  We are able
to detect the power on large scales because the Kast spectra and the
continua that we fit are more stable over these large scales than are
high resolution echelle spectra and their continua \citep{suzuki03b}.
The \sigzo\ that we measure is larger than a value we calculated, for
no clear reason. Both the measurement and calculation could have large errors.

The dispersion of DA measured in segments of spectra 121.567~\AA\ long
in the observed frame at $z=1.9$ for \lya\ comes about equally from
the low density IGM, LLS and metal lines.

The flux field is significantly different from a random Gaussian
field, with an enhanced probability of a large amount of absorption,
on all scales 10 -- 10,000~\kms .  On small scales the asymmetry comes
from the density distribution in the low density IGM, making spectra
that are largely absorption free, with occasional \lya\ lines.  On
large scales the asymmetry comes from the rare high density regions
that make absorption with large H~I column densities. They make LLS
and DLAs with strong \lya\ lines, and they place many strong metal
lines all across a spectrum.

\subsection{Comparison with Hydrodynamic Simulations}

We find that a hydrodynamic simulation on a $1024^3$ grid in a
75.7~Mpc box reproduces the observed mean DA from the IGM alone when
we use popular parameters H$\rm _o=71$~\kms Mpc$^{-1}$, $\Omega_b =
0.044$, $\Omega_m = 0.23$ and ${\Omega_{\Lambda} = 0.73}$, ${\sigma_8
= 0.9}$ and a UV background with an ionization rate per H~I atom of
$\Gamma _{912}= (1.44 \pm 0.11) \times 10^{-12}$~s$^{-1}$ that is $
\gamma _{912} =1.08 \pm 0.08$ times the value calculated by
\citet{madau99} with 61\% from QSOs and 39\% from stars. This value of
the $\Gamma $ is similar to the $\Gamma _{912}> 1.5 \times 10^{-12}$
s$^{-1}$ from \citet{steidel01} at $z \simeq 3$.

The mean DA that we measure for the \lya\ from the low density IGM
provides a joint constraint on two cosmological parameters, \ob\ and
$\sigma_8$, and one astrophysical parameter, $\Gamma _{912}$.  Using
\ob $=0.0444 \pm 0.0018 $, $\sigma _8 = 0.9 \pm 0.1$ and DA $= 0.118
\pm 0.010$ we find \ $\Gamma _{912}= (1.44 \pm 0.36) \times
10^{-12}$~s$^{-1}$, where the error includes the contributions from the
errors in \ob , $\sigma _8$ and DA, but not \om\ or \ol .

The baryon density that accounts for the DA in the IGM at $z=1.9$ is
the same as the value measured using D/H and the CMB, with an
uncertainty of about 6\%.  When we use $\sigma _8 = 0.9 \pm 0.1$,
$\Gamma = 1.329 \pm 0.133 \times 10^{-12}$~s$^{-1}$ and DA $= 0.118 \pm 0.010$,
we found \ob $= 0.0417 \pm 0.0022$, where the error is from the DA
alone (Table 5), or $\pm 0.0085$ using the errors on all the listed
parameters. This value agrees with the D/H + BBN value, \ob $=0.042
\pm 0.004$ \citep{kirkman03}, and with the value from the first year
WMAP data, \ob $= 0.0444 \pm 0.0018$ \citep{spergel03}.

When we pass multiple lines of sight through the hydrodynamic
simulation, we see slightly less variation in the DA in $\Delta z =
0.1$ than the Kast spectra IGM only, in part because of the box lacks
large scale power.  The power of the flux in these lines of sight does
match that in HIRES spectra on scales $0.008 < \log k < 0.08$ (s/km).

\subsection{Opportunities for Improvement}

The values in Table \ref{taba} suggest that we could improve the
accuracy of our mean DA4 measurement by observing many more QSOs, even
without any improvements in the methods. The measurement error for the
Kast spectra (column 4) would approach that from the S/N and continuum
fits (column 3) with a sample about 4 times larger, or 300 QSOs.  Here
we assume that the DA4 values, each averaged over a path of 19.75,
will be nearly normally distributed.  The values given here are
approximate, since they assume that the artificial spectra are an
adequate representation of the IGM, even though they do not explicitly
include absorption from high \nhi\ lines and metal lines.

We find that S/N = 6 per 1.13~\AA\ in the observed frame is adequate
for continuum placement in Kast spectra.  With improved continuum
placement methods we might be able to use lower S/N, but we would not
be able to adjust the continua to fit the emission lines of each
individual QSO, and hence the continuum fit errors will complicate the
measurement of the large scale power.

High resolution spectrographs have the major advantage of allowing us
to find and remove the individual \lya\ lines of LLS and the metal
lines in the \lyaf . This reduces the sample size required for a given
$\sigma (DA)$ by about a factor of four, and it greatly improves the
accuracy of the corrections for the LLS and metal lines. The factor of
four largely compensates for the lower efficiency of high resolution
spectrographs in terms of photon recorded per \AA\ per second.
However, to find metal lines in high resolution spectra we would
prefer S/N $>10$ per 0.03~\AA , which is about 100 times more photons
per \AA\ than we have with Kast spectra.  We would also require
improvement in the flux calibration and continuum fits to the high
resolution spectra \citep{suzuki03b}.

\section{Acknowledgments}


This work was funded in part by grant NAG5-13113 from NASA and by
grant AST-0098731 form the NSF.  The spectra were obtained from the
Lick observatory and we thank the Lick Observatory staff.  We are very
grateful to Avery Meiksin and Rupert Croft for drawing our attention
to this topic, and helping us understand the issues.  We are
especially grateful to Pat McDonald for providing us with the
artificial absorption spectra and their description, and for many detailed
comments on the manuscript.  This research has made extensive use of
the NASA/IPAC Extragalactic Database (NED) which is operated by JPL,
under contract with NASA.

\clearpage

\bibliographystyle{apj}
\bibliography{archive}

\begin{deluxetable}{cccrc}
\tablecaption{\label{tabd} DA as a Function of Redshift}
\tablewidth{0pt}
\tablehead{
\colhead{$z$} &
\colhead{Mean DA} &
\colhead{$\sigma (\mu )$} &
\colhead{15.8\% below} &
\colhead{15.8\% above}
}
\startdata
1.64  &  0.1124  &  0.0089  &  -0.1303   &  0.3410\\
1.72  &  0.1359  &  0.0041  &  -0.0436   &  0.3160\\
1.80  &  0.1333  &  0.0034  &  -0.0143   &  0.2910\\
1.88  &  0.1398  &  0.0028  &   0.0015   &  0.2955\\
1.96  &  0.1681  &  0.0030  &   0.0071   &  0.3492\\
2.04  &  0.1714  &  0.0033  &   0.0114   &  0.3526\\
2.12  &  0.1973  &  0.0047  &   0.0168   &  0.4018\\
2.20  &  0.1775  &  0.0052  &   0.0185   &  0.3570\\
2.28  &  0.1667  &  0.0113  &   0.0001   &  0.3522\\
2.36  &  0.1576  &  0.0332  &  -0.0874   &  0.4461
\enddata
\end{deluxetable}
\newpage 

\begin{deluxetable}{l|lcccl}
\tablecaption{\label{tabc} Absorption by Metal Lines}
\tablewidth{0pt}
\tablehead{
\colhead{Parameter} &
\colhead{$\lambda  _r $ (\AA )} &
\colhead{$\lambda  _{obs} $ (\AA )} &
\colhead{$\sigma $ (\%) } &
\colhead{Mean (\%) } &
\colhead{$\sigma (Mean) $ (\%) } 
}
\startdata
DM1        & 1362   & ...  & ... & 1.58 & 0.13\\
DM2        & ...    & 4158 & ... & 1.58 & 0.13\\
DM3        & 1360   & 4135 & 2.5 & 1.87 & 0.13\\
DM4 SBS    & 1358   & 4125 & 2.7 & 1.67 & 0.22\\ 
DM5 SBS    & 1360   &  ... & ... & 1.66 & 0.22\\ 
DM6 SBS    &  ...   & 4158 & ... & 1.65 & 0.22\\ 
DM7(CIV)   & ...    & 4580 & ... & 0.69 & 0.15\\
DM7(CIV)   & 1360   & ...  & ... & 0.71 & ...\\
DM7(CIV)   & 1120   & ...  & ... & 0.80 & ...\\
DM1-6      & 1120   & ...  & ... & 1.92 & 0.42\\ 
DMs        & 1120   & ...  & ... & 2.3 & 0.5\\
\enddata
\end{deluxetable}
\newpage 

\begin{deluxetable}{l|cclllll}
\tablecaption{\label{tabb} Distribution of the DA2s 
                           at $z=1.9$ in $\Delta z = 0.1$ Segments}
\tablewidth{0pt}
\tablehead{
\colhead{Data Set} &
\colhead{Metals?} &
\colhead{LLS?} &
\colhead{n} &
\colhead{$\sigma $(DA2)} & 
\colhead{$\sigma [\sigma$(DA2)}] & 
\colhead{Mean DA2} &
\colhead{$\sigma {\mu}$}
}
\startdata
Kast Spectra          & yes & yes &  152 & 0.0612 & 0.0035 & 0.1563 & 0.0050\\
Kast $< 0.35$         & yes & yes &  151 & 0.0487 & 0.0028 & 0.1488 & 0.0049\\  
Ly-$\alpha $ of LLS   & no &  yes & 471: & 0.0351 & 0.0047 & 0.0103 & 0.004\\
Metals (DM3, DMs)     & yes & no  & 377  & 0.031  & ...    & 0.023  & 0.005\\
\sigzo Kast IGM only  & no & no  &  152 & 0.039 & $^{+0.005}_{-0.007}$
& 0.118  & 0.010\\  
Artificial Spectra    & no &  no  &  150 & 0.0422 & 0.0024 & 0.1518 & 0.0035\\
Hydro. Spectra        & no & no  &  150 & 0.0331 & 0.0019 & 0.1287 & 0.0027
\enddata
\end{deluxetable}
\begin{deluxetable}{c|ccccc}
\tablecaption{\label{taba} DA3s Values for Quarters of the Kast and Artificial Spectra}
\tablewidth{0pt}
\tablehead{
\colhead{} &
\colhead{Artificial} &
\colhead{Artificial} &
\colhead{Artificial True -} &
\colhead{Kast} &
\colhead{Mean}
\\
\colhead{} &
\colhead{True} &
\colhead{Measured} &
\colhead{Measured} &
\colhead{Measured} &
\colhead{$z_{abs}$}
}
\startdata
First Quarter   & 0.1574 &  0.1577  & -0.00027  &   0.1487 &  1.949 \\
Second Quarter  & 0.1668 &  0.1660  &  0.00075  &   0.1456 &  1.936 \\
Third Quarter   & 0.1433 &  0.1486  & -0.00531  &   0.1696 &  1.966 \\
Fourth Quarter  & 0.1546 &  0.1449  &  0.00990  &   0.1404 &  1.846 \\
\hline 
$\sigma$        & 0.0097 &  0.0095  &  0.00634  &   0.0128 &   ...  \\
Mean $\mu $     & 0.1555 &  0.1543  &  0.00127  &   0.1511 &  1.924 \\
$\sigma (\mu )$ & 0.0048 &  0.0048  &  0.00317  &   0.0064 &   ...  \\
\enddata
\end{deluxetable}
\newpage 

\begin{deluxetable}{cccccc}
\tablecaption{\label{tabe} DA8s for Various combinations of 
                           Cosmological Parameters}
\tablewidth{0pt}
\tablehead{
\colhead{Parameter} &
\colhead{$\Omega _b$} &
\colhead{$\sigma _8 $ } &
\colhead{$\gamma _{912}$} &
\colhead{$\gamma _{228}$} &
\colhead{DA8s}\\
\colhead{measured} &
\colhead{} &
\colhead{} &
\colhead{} &
\colhead{} &
\colhead{}
}
\startdata
all & $0.0444 \pm 0.0018 $ & $0.9 \pm 0.1$ & $1.0 \pm 0.1$ & 1.8 & $0.118 \pm 
0.010$\\
DA in sim. & 0.0440 & 0.9 & 1.0 & 1.8 & $0.1287 \pm 0.0027$\\
$\gamma _{912}$ & 0.0440 & 0.9 & $1.08 \pm 0.08$ & 1.8 & 0.118\\
$\sigma _8$ & 0.0440 & $0.94 \pm 0.04$ & 1.0 & 1.8 & 0.118\\
$\Omega _b $ & $0.0417 \pm 0.0022$ & 0.9 & 1.0 & 1.8 & 0.118\\
$\gamma _{912}$ & $0.0444 \pm 0.0018 $ & $0.9 \pm 0.1$ & $1.08 \pm 0.27$ 
& 1.8 & $0.118 \pm 0.010$\\
\enddata
\end{deluxetable}


\clearpage

\end{document}